\documentclass[lettersize,journal]{IEEEtran}
\usepackage{amsmath,amsfonts}
\usepackage{amssymb}
\usepackage{algorithmic}
\usepackage[ ruled, vlined, linesnumbered]{algorithm2e}
\usepackage{array}
\usepackage{textcomp}
\usepackage{stfloats}
\usepackage{url}
\usepackage{verbatim}
\usepackage{graphicx}
\usepackage{cite}
\usepackage{caption}
\usepackage{subcaption}
\begin{document}

\title{A Novel Defense Against Poisoning Attacks on Federated Learning: LayerCAM Augmented with Autoencoder}

\author{Jingjing Zheng,~\IEEEmembership{Graduate Student Member,~IEEE,} Xin Yuan,~\IEEEmembership{Member,~IEEE,} Kai Li,~\IEEEmembership{Senior Member,~IEEE,}  Wei Ni,~\IEEEmembership{Fellow,~IEEE,} Eduardo Tovar,~\IEEEmembership{Member,~IEEE,} and Jon Crowcroft,~\IEEEmembership{Fellow,~IEEE,}
\thanks{J. Zheng is with CyLab Security and Privacy Institute, Carnegie Mellon University, Pittsburgh, PA 15213, USA, and also with Real-Time and Embedded Computing Systems Research Centre (CISTER), Porto 4249–015, Portugal (E-mail: zheng@isep.ipp.pt).}

\thanks{X. Yuan is with the Digital Productivity and
Services Flagship, Commonwealth Scientific and Industrial Research Organization (CSIRO), Sydney, NSW 2122, Australia (E-mail: xin.yuan@data61.csiro.au).}
        
\thanks{K. Li is with the Department of Engineering, University of Cambridge, CB3 0FA Cambridge, U.K., and also with Real-Time and Embedded Computing Systems Research Centre (CISTER), Porto 4249–015, Portugal (E-mail: kaili@ieee.org).}

\thanks{W. Ni is with the Digital Productivity and
Services Flagship, Commonwealth Scientific and Industrial Research Organization (CSIRO), Sydney, NSW 2122, Australia (E-mail: wei.ni@data61.csiro.au).}

\thanks{E. Tovar is with Real-Time and Embedded Computing Systems Research Centre (CISTER), Porto 4249–015, Portugal (E-mail: emt@isep.ipp.pt).}

\thanks{Jon Crowcroft is with the University of Cambridge, Cambridge CB2 1TN, U.K., and also with the Alan Turing Institute, London NW1 2DB, U.K. (e-mail: jon.crowcroft@cl.cam.ac.uk).}
\thanks{Manuscript received May 31, 2024}}

\markboth{Journal of \LaTeX\ Class Files,~Vol.~14, No.~8, June~2024}%
{Shell \MakeLowercase{\textit{et al.}}: A Sample Article Using IEEEtran.cls for IEEE Journals}


\maketitle

\begin{abstract}
Recent attacks on federated learning (FL) can introduce malicious model updates that circumvent widely adopted Euclidean distance-based detection methods. This paper proposes a novel defense strategy, referred to as LayerCAM-AE, designed to counteract model poisoning in federated learning. The LayerCAM-AE puts forth a new Layer Class Activation Mapping (LayerCAM) integrated with an autoencoder (AE), significantly enhancing detection capabilities. Specifically, LayerCAM-AE generates a heat map for each local model update, which is then transformed into a more compact visual format. The autoencoder is designed to process the LayerCAM heat maps from the local model updates, improving their distinctiveness and thereby increasing the accuracy in spotting anomalous maps and malicious local models. To address the risk of misclassifications with LayerCAM-AE, a voting algorithm is developed, where a local model update is flagged as malicious if its heat maps are consistently suspicious over several rounds of communication. Extensive tests of LayerCAM-AE on the SVHN and CIFAR-100 datasets are performed under both Independent and Identically Distributed (IID) and non-IID settings in comparison with existing ResNet-50 and REGNETY-800MF defense models. Experimental results show that LayerCAM-AE increases detection rates (Recall: 1.0, Precision: 1.0, FPR: 0.0, Accuracy: 1.0, F1 score: 1.0, AUC: 1.0) and test accuracy in FL,  surpassing the performance of both the ResNet-50 and REGNETY-800MF.  Our code is available at: https://github.com/jjzgeeks/LayerCAM-AE
\end{abstract}


\section{Introduction}

Federated learning (FL) has recently emerged as a promising distributed machine learning paradigm where multiple users collaboratively train a shared machine learning model under the orchestration of a server, while retaining the training data on individual users to mitigate the risk of their privacy data leakage \cite{zhou2022}. In FL, the users, such as hospitals, are equipped with storage and computation capability and can locally hold and process the patients' data, e.g., medical imaging datasets or clinical records \cite{kaissis2021end}. The users iteratively train their local models (i.e., weight parameters or gradients) by taking advantage of their private data, and send the local models instead of the raw private data to a server for aggregation.  The server utilizes the local models to compute a global model that is sent back to the users for updating their local models \cite{mcmahan17a}.  This process is repeated in multiple communication rounds until a desirable accuracy of image classification is achieved or convergence, and it significantly alleviates the leakage of private data from the source.

While FL is able to enhance user data privacy protection, the distributed nature in FL is insufficient to ensure that every local model update is benign. An attacker-controlled malicious user can potentially launch model poisoning attacks by directly manipulating local model parameters and propagating it to the global model, resulting in a corruption of the FL global model \cite{fang2020local, Cao_2022_CVPR}. To defend against model poisoning attacks in FL, existing distance-based defense mechanisms, e.g., Euclidean distance \cite{blanchard2017machine, yin2018byzantine} or cosine similarity \cite{CaoF0G21}, have been developed to filter out suspicious or unreliable local models before the server aggregation. Nevertheless, these defense measures confront the following issues. On the one hand, excessive deletion of local model updates and/or expensive analysis of high-dimensional local model updates incur inefficiency and degradation of model quality to a large extent. On the other hand, the attacker, by eavesdropping on the benign local models to craft a malicious local model update that closely resembles benign local models, which can circumvent existing defense countermeasures such as Krum \cite{blanchard2017machine}, Trim-mean \cite{yin2018byzantine}, and Geometric Median \cite{geometricmedian}.


In this paper, we propose a novel malicious-user detection method called LayerCAM-AE that identifies and filters malicious local model updates of FL from the server side. Since Layer Class Activation Mapping (LayerCAM) can visualize and interpret the off-the-shelf deep neural networks (DNNs) by highlighting the regions of an input image or feature map that contribute the most to the model's predictions, LayerCAM is leveraged by LayerCAM-AE to generate discriminate a heat map for every local model update before server aggregation. LayerCAM-AE not only cleverly maps the high-dimensional local model update into a low-dimensional, reliable and precise fine-grained heat map but also indirectly visualizes the local model update. 

To further eliminate the potential errors of LayerCAM-assisted malicious user identification, i.e., a genius attacker utilizes Graph Autoencoder (GAE) to reconstruct a malicious local model that can capture the correlation of benign local model updates  \cite{li2024data} and may evade state-of-the-art defense mechanisms,  autoencoder is embedded in LayerCAM-AE to remap the LayerCAM heat maps to accentuate the hidden features of the heat maps, and improve the distinguishability of the heat maps and the success rate of discerning anomalous heat maps and malicious local model updates. 

The main contributions of this paper are summarized as follows.
 

\begin{itemize}
\item We propose a new defense strategy for model poisoning attacks on FL, leveraging extended LayerCAM and autoencoder to effectively detect inconspicuous manipulations.

\item The new LayerCAM is developed to transform the high-dimensional, indistinct local model updates in FL into low-dimensional, visually interpretable heat maps.

\item The autoencoder is incorporated into the proposed LayerCAM-AE, refining the LayerCAM-generated heat maps to highlight their latent features and enhance their distinguishability. This improvement aids in more effectively identifying anomalous heat maps and detecting malicious local models.

\item In addition, we proposed a voting algorithm to consistently filter out transient malicious model updates, thereby reducing the likelihood of erroneously identifying malicious local models.
\end{itemize}

We conducted a comprehensive assessment of the proposed LayerCAM-AE framework using two public datasets, SVHN and CIFAR-100, under both IID and non-IID settings. Our assessment encompasses two prominent deep learning models, i.e., ResNet-50 \cite{ResNet} and REGNETY-800MF \cite{regnet}. Our approach offers superior detection rates and FL test accuracy compared to the state-of-the-art methods.

\section{Federated Learning}
In this section, we exemplify image classification to delineate the training process of FL.
The system is composed of a server and $N$ benign users where $\mathcal{N} \in  \triangleq \{1, \cdots, n \cdots, N \}$ and $n$ is the index of the user. Each benign user trains a Deep Neural Network (DNN) that is particularly well-suited for image classification. The DNN analyzes an input image (e.g., cup), extracts features (e.g., shape, handle, size and rim) through several convolutional (CONV)  layers, feeds the feature maps into fully connected (FC) layers, and assigns one of the labels (such as ``bus'', ``cup'', ``fox'' or ``boy''), see as Figure \ref{fl-layer-cam}. The DNN model is learned based on the training data and is represented by the parameter vector  $\mathbf{W}$ (weights and biases). The user owns  $D_n$ pairs of training data $\mathcal{D}_n = \{ (\mathbf{x}_d, y_d) \}_{d = 1}^{D_{n}}$, consisting of the feature vector  $\mathbf{x}_s$ as input to the model (e.g., pixels of an image) and the corresponding scalar value $y_d$ (e.g., the real label of the image), which is the output required by the model the desired output of the model \cite{shiqiangwang2019}. 

Let $f_d(\mathbf{W}, \mathbf{x}_d, y_d)$ denote the loss function of each training data sample $d$, which captures approximation errors over the input  $\mathbf{x}_d$ and desired output $y_d$. $\mathbf{W}$ is the weight parameter of the loss function in the neural network being trained according to the FL procedure. $f_d(\mathbf{W}, \mathbf{x}_d, y_d)$ can be specified according to the machine learning models, for instance, $f_d(\mathbf{W}, \mathbf{x}_d, y_d) = \frac{1}{2}\big( y_d-\mathbf{W}^T\mathbf{x}_d\big)^2 $ is used to model linear regression, or $f_d(\mathbf{W}, \mathbf{x}_d, y_d) = - y_d \log\big(\frac{1}{1 + \exp(-\mathbf{W}^T\mathbf{x}_d)}\big) - (1 - y_d) \log\big(1 - \frac{1}{1 + \exp(-\mathbf{W^T}\mathbf{x}_d)}\big)$ is used to model logistic regression. Here, $\mathbf{W}^T$ is the transpose matrix of $\mathbf{W}$). For each user $n$, given the dataset $\mathcal{D}_{n}$,  the local loss function of the parameter vector $\mathbf{W}$ on the collection of data samples is defined as 
     \begin{equation}
       F_n (\mathbf{W}) \triangleq \frac{1}{D_{n}} \sum_{d = 1}^{\mathbf{D}_{n}} f_d(\mathbf{W}, \mathbf{x}_d, y_d).
       \label{local-loss}
     \end{equation}

Accordingly, the global loss function on all the distributed datasets evaluated on parameter vector $\mathbf{W}$ is given by
\begin{equation}
    F(\mathbf{W}) = \sum_{n = 1 }^N \frac{D_n}{D}   F_n (\mathbf{W}), 
       \label{global-loss}
\end{equation}
where $D =  \sum_{n = 1}^N D_n$. The goal of the training process is  to find optimal parameters $\mathbf{W}^{*}$ such that the global loss function $F(\mathbf{W})$ is minimized, i.e.,
      \begin{equation}
          \mathbf{W}^{*} = \arg \mathop{\min} \limits_{\mathbf{W}} F(\mathbf{W}).
          \label{optimal-weight}
      \end{equation}
To solve the problem in \eqref{optimal-weight} while preserving the data privacy for each device, a canonical gradient-descent technique is widely used in state-of-the-art FL systems \cite{mcmahan17a}, which is implemented iteratively in a distributed manner as follows.

Let $T_{FL}$ denote the total number of communication rounds with $\mathcal{T} \triangleq \{0, \cdots,  T_{FL} -1\}$, one communication round refers to users download global model parameter from server and send their update local model parameters to server. At each communication round $t \in \mathcal{T}$, the process of the FL system contains the following three steps:

\begin{itemize}
    \item \textit{User selection and broadcast:} The server samples $N$ users (without loss of generality, we assume that all $N$ users participate in FL training) satisfying eligibility requirements. The selected users download the current global model $\mathbf{W}^t$ from the server.
    
    \item \textit{Local model training and uploading:} Each user $n$ randomly samples mini-batch, i.e., $\mathcal{\tilde{D}}_n $ ($ \mathcal{\tilde{D}}_n \in \mathcal{D}_n$ ),  from the local dataset $\mathcal{D}_n$ and utilizes the downloaded global model $\mathbf{W}^t$ to perform a total of $L$ local training epochs to update the local model, i.e., 
    \begin{equation}
       \mathbf{W}_{n, \ell+1}^t =  \mathbf{W}_{n, \ell}^t  - \eta_t \nabla  F_n( \mathbf{W}_{n, \ell}^t, \mathcal{\tilde{D}}_n), ~ \ell \in \{ 0, 1,\cdots, L-1 \},
       \label{local-model}
    \end{equation}
    where $\eta_t$ denotes the learning rate at communication round $t$, $\nabla  F_n( \mathbf{W}_{n, \ell}^t, \mathcal{\tilde{D}}_n)$ defines the local
gradient estimate over $\mathcal{\tilde{D}}_n$ at local training epoch $\ell$. After $L$ local training epochs, each user $n$ uploads the local model updates $\mathbf{W}_n^{t} = \mathbf{W}_{n, L}^t$ to the server. 

 \item  \textit{Global model aggregation and update:} The server aggregates $N$ participants’ updated models $\{ \mathbf{W}_{n}^{t} \}_{n=1}^N$  to obtain the updated global model as
\begin{equation}
    \mathbf{W}^{t+1} =  \sum_{n =1}^N \frac{D_n}{D} \mathbf{W}_{n}^{t},
    \label{global-model}
\end{equation}
where $ \mathbf{W}^{t+1}$ is sent back to all $N$ participants. 
\end{itemize}
The above process repeats until the number of communication rounds $T_{FL}$ is met.

\section{Poisoning Attacks Against FL}
Model poisoning attack is a striking security threat in FL,  where malicious users try to compromise the global model by injecting poisoned data into the local model or deliberately tampering with local model parameters during the training process. In our attack settings, the model poisoning attack launched by the attackers whose goal is to degrade or impair the FL performance, such as accuracy and convergence, eventually resulting in denial-of-service attacks. 

During the process, all benign local models are uploaded to the server, and the attackers with $\mathcal{M} \triangleq \{1', \cdots, k',  \cdots, K' \} $ keep the same DNN structure as benign users.
Unaware of the ill-intentioned attacker, the server collects the local model updates of all users, including both the benign and malicious ones, and unconsciously creates a
contaminated global model update,  denoted by $\mathbf{W}_g^{t+1}$,  at the $t$-th
communication round, i.e., 
\begin{equation}
	\mathbf{W}_g^{t+1} =  \sum_{n = 1 }^N \frac{D_n}{S} \mathbf{W}_{n}^{t} + 
	\frac{D_{k'}}{S}  \mathbf{W}_{k'}^{t},
	\label{aggregate-models}
\end{equation}
where $S = \sum_{n = 1 }^N D_n +  D_{k'} $ is the total size of the local training data reported to the server, and $D_{k'}$ is the claimed data size of the attacker $k'$.

In case the server is unaware that the benign local models and malicious local models are mixed, the server unintentionally minimizes the following global loss function:
\begin{equation}
\begin{split}
    \mathop{\min}_{\mathbf{W}_g^{t+1}}   F (\mathbf{W}_g^{t+1})   = \sum_{n= 1}^N \frac{D_n}{S}   F_n (\mathbf{W}_g^{t+1}) + \frac{D_{k'}}{S}   F_{k'} (\mathbf{W}_g^{t+1}), 
 \end{split}
 \label{contaminated-global-model}
\end{equation}
where $ F_{k'}(\cdot)$ is the local loss function of the attacker, which conforms to (\ref{local-loss}).

To impair the FL training model, the attackers aim to maximize the FL global loss function $F (\mathbf{W}_g^{t+1})$, i.e., equivalent to minimizing the training accuracy of FL, while keeping $ \mathbf{W}_{k'}^{t}$ imperceptible by the server that Euclidean distance-based \cite{blanchard2017machine, yin2018byzantine} or similarity-based detection \cite{CaoF0G21}. In other words,  the attacker can measure the Euclidean distance between $\mathbf{W}_{k'}^{t}$ and $\mathbf{W}_g^{t+1}$,  which is ensured to be smaller than a threshold $r$. Mathematically formalized as follows, 
\begin{subequations}\label{attacker-obj2}
	\begin{align}
		&  \mathop{\max}_{\mathbf{W}_{k'}^{t}} F(\mathbf{W}_g^{t+1}) 
		\label{attacker-obj}\\
		&\qquad \text{s.t} ~ \mathfrak{D} \big(\mathbf{W}_{k'}^{t}, \mathbf{W}_g^{t+1} \big) \leq r, \forall k' \in \mathcal{M},
		\label{attacker-constraint}
	\end{align}
\end{subequations}
where $\mathfrak{D} (\cdot)$ is the Euclidean distance function $\mathfrak{D} \big(\mathbf{W}_{k'}^{t}, \mathbf{W}_g^{t+1} \big) = \| \mathbf{W}_{k'}^{t} -  \mathbf{W}_g^{t+1} \|_2$, and $r$ is a predetermined threshold
that ensures the created malicious local model update is approximate to the global model within the Euclidean distance domain to bypass existing Euclidean-based distance defense mechanisms on the server side.

Note that the solution to optimization \eqref{attacker-obj2} can be found in the latest literature \cite{li2024data}, and the goal of this paper is not to solve this optimization problem. We are designed to defend against such attacks.

\section{Proposed FL-LayerCAM against FL Poisoning Attacks}
The malicious local models are mixed with benign local models
and uploaded to the server by attackers. The set of benign users and attackers is $\mathcal{P} = \mathcal{N} \cup  \mathcal{M} $ with cardinality $|\mathcal{P}| =  N + K'$.

\subsection{LayerCAM-AE Architecture}
\begin{figure*}[htbp]
  \centering
  \includegraphics[width=1.0\textwidth]{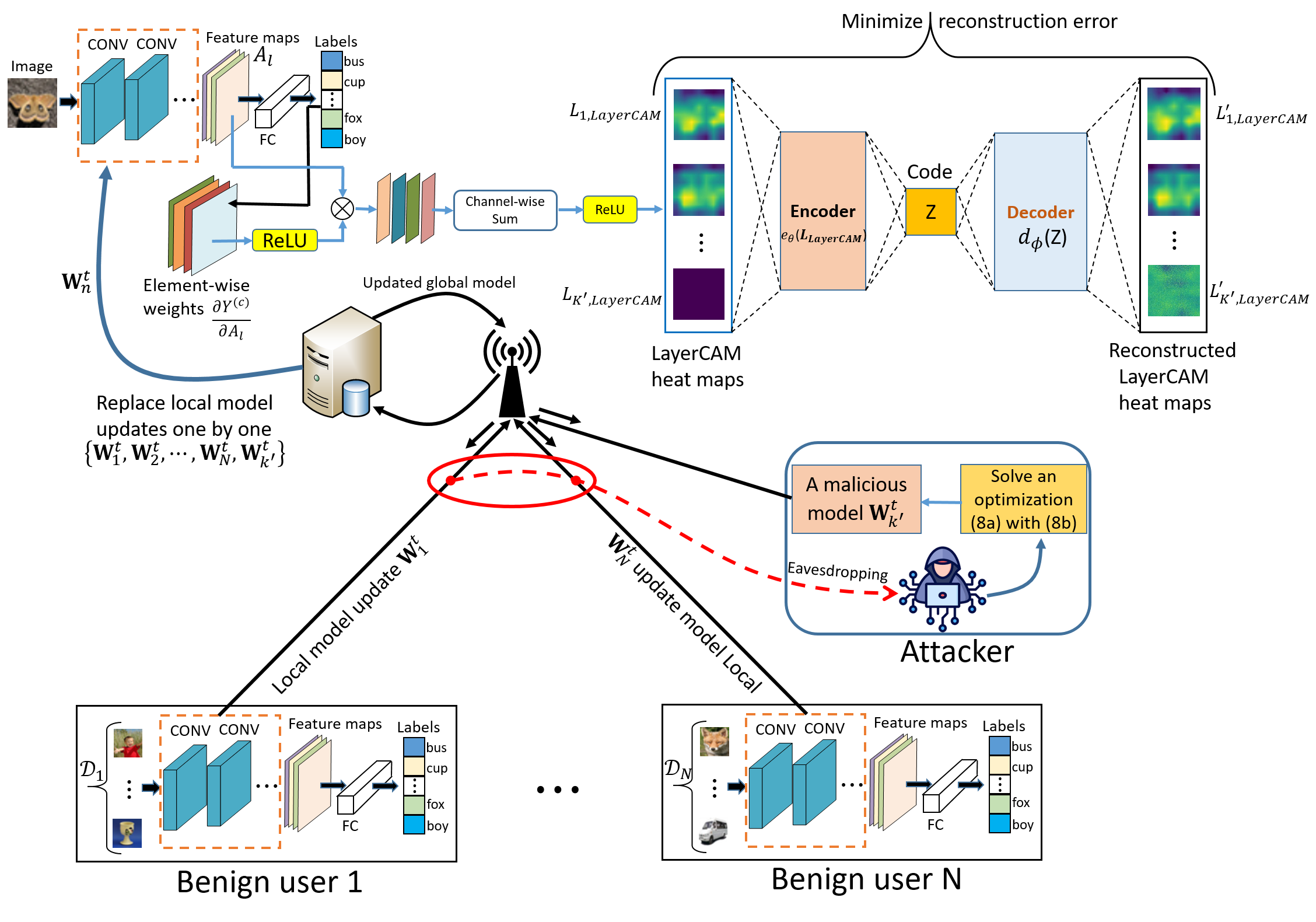}
  \caption{Illustration of the proposed LayerCAM-AE framework, where 
  the server arbitrarily selects an image (e.g., an image with the label ``butterfly'') from the global model testing dataset to create LayerCAM heat maps for every uploaded model update. These LayerCAM heat maps flow into an autoencoder. Large reconstruction errors of heat maps are judged malicious. }
\label{fl-layer-cam}
\end{figure*}

On the user side, each of the $K$ benign users utilizes the DNN tailored specifically for image classification tasks, as shown in Figure~\ref{fl-layer-cam}. The DNN model extracts relevant features from an input image (e.g., a butterfly) and subsequently maps them to the corresponding classes. The architecture of a typical DNN comprises multiple layers, each with a specific function. 
$(a) $
The first layer is a convolutional layer, which applies a set of filters to the input image, thereby extracting intrinsic features,  such as legs, thorax, and abdomen. The output of the convolutional layer is a set of feature maps, each representing different aspects of the image. 
$(b)$
The following layer is a pooling layer, which reduces the dimensionality of the feature maps while retaining essential information, such as the salient features of the butterfly wings. Various pooling methods can be employed, including max pooling or average pooling, all with the objective of downsampling the feature maps while retaining their salient features. Some DNN architectures may differ from traditional pooling operations. For example, models like SqueezeNet \cite{SqueezeNet}, ResNet \cite{ResNet}, DenseNet \cite{DensNet}, and MobileNet \cite{Howard_2019_ICCV} employ alternative strategies.
$(c)$
After several convolution and pooling layers, the extracted features are fed into one or multiple fully connected layers, where the final classification is performed. 

Upon receiving the local model updates from the users, the server aggregates the local models, where the benign local models can be mingled with malicious local models. We aim to design a defense countermeasure that can identify and remove malicious users. LayerCAM \cite{layercam} can produce reliable and high-quality class activation maps for different layers of DNN,  and improves the localization accuracy of activated regions compared to CAM and GradCAM. By using multi-scale features and hierarchical representations, LayerCAM can better localize the discriminative regions in the image that contribute to the model's prediction compared to other methods, e.g., Class Activation Mapping \cite{Zhou_2016_CVPR},  Grad-CAM \cite{selvaraju2017grad} and  Grad-CAM++ \cite{Grad-CAM++} \cite{selvaraju2017grad}. Unlike GradCAM, which may suffer from reduced performance with very deep networks due to vanishing gradients, LayerCAM is less affected by network depth. It achieves this by aggregating activation maps from multiple layers, ensuring that informative features are captured at each level of the network~\cite{layercam}.



Figure \ref{fl-layer-cam} illustrates our proposed LayerCAM-AE architecture that integrates LayerCAM and autoencoder, where LayerCAM is leveraged to generate heat maps for all local model updates of FL, and autoencoder is utilized to identify abnormal heat maps. Server identification of malicious local model updates consists of two components: LayerCAM-based process and autoencoder-based process.

\textbf{\underline{LayerCAM-based process}}. The server randomly picks an image from the global model testing dataset that incorporates all categories of the users' dataset as input and passes through the convolutional layers with weight and bias parameters that are replaced by each model update $\mathbf{W}_l,  l \in \mathcal{K} \cup \mathcal{M} $ ($l$ is the index of local model updates, including benign and malicious), obtains the feature maps\footnote{In this paper, we focus on extracting the feature maps after the final convolution layer of deep learning model in that can capture high-level features and hold information regarding the important regions of the input image.} $A_l$ that has $K$ channels.  The feature maps $A_l$ are fed into a fully connected layer for final classification. To obtain the class discriminative localization map of $L_{l,LayerCAM}^{(c)} \in R^{B \times H} $ with width $B$ and height $H$  for any class $c$, LayerCAM first computes the gradient of the score for class $c$, $Y^{(c)}$ (before softmax) with respect to each spatial location $(i, j)$ in the $k$-th feature map within $A_l$, i.e., $\frac{\partial Y^{(c)}}{\partial A_l^k(i,j)}$, $k \in [1, K]$  is the index of channels. The importance of element-wise weights $\alpha_{l,ij}^{(c)}$ can be obtained through the rectified linear unit (ReLU):
\begin{equation}
	\alpha_{l,ij}^{(c)} =  ReLU  \Big( \frac{\partial Y^{(c)}}{\partial A_{l}^m(i, j)} \Big), \forall l \in \mathcal{K} \cup \mathcal{M},
	\label{layer-importance}
\end{equation}
where $A_{l}^m(i, j)$ is the activation at spatial location $(i, j)$ of the feature map $A_l^k$. LayerCAM further multiplies the activation value of each location in the feature map by an importance weight, the multiplications are linearly combined along the channel dimension to obtain the class activation map, and the result is through ReLu operation, which is formulated as follows:
\begin{equation}
	L_{l,LayerCAM}^{(c)} = ReLU  \Big( \sum_{m = 1}^n \alpha_{l,ij}^{{(c)}} \cdot A_{l}^k(i, j)  \Big).
	\label{LayerCAM-maps}
\end{equation}

\textbf{\underline{Autoencoder-based process}}. Each LayerCAM heat map  $L_{l, GradCAM}$ with size of $H \times B$ generated from a test image and a local model update is flattened into a vector with size of $1 \times H \times B$, which is further concatenated with the vectors of other LayerCAM heat maps to form $\mathbf{L_{LayerCAM}}$ as the input to the encoder. During the autoencoder training, the encoder $e_{\theta}(\mathbf{L_{LayerCAM}})$ with parameter $\theta$ compresses the LayerCAM heat maps from a high-dimensional space to a low-dimensional space, i.e., $z = e_{\theta}(\mathbf{L_{LayerCAM}})$, also called the code or the latent space. The code learns the underlying features or representation of the LayerCAM heat maps, which are input to the decoder $d_{\phi}(z)$ with parameter $\phi$. The decoder further reconstructs the input LayerCAM heat maps from the code, i.e.,  $d_{\phi}(z) = \mathbf{L_{{LayerCAM}}^{'}}  =  d_{\phi}\big(e_{\theta}(\mathbf{L_{LayerCAM}}) \big)$. After the training of $(\theta, \phi)$, the reconstructed LayerCAM heat maps are reshaped into the same size as the original LayerCAM heat maps.  To minimize the difference between the original input LayerCAM heat maps and reconstructed LayerCAM heat maps, the autoencoder loss function is defined as
the mean squared error (MSE) between the encoder input
LayerCAM heat maps $\mathbf{L_{LayerCAM}}$, and the decoder 
reconstructed LayerCAM heat maps $\mathbf{L^{'}_{LayerCAM}}$, i.e.,
\begin{equation}
	\begin{split}
		L(\theta, &\phi)  = \mathop{\min}_{\theta, \phi} \frac{1}{|N| + |K'|} \| \mathbf{L_{LayerCAM}} - \mathbf{L^{'}_{LayerCAM}} \|^2_2 \\
		= & \!\mathop{\min}_{\theta, \phi}  \!\frac{1}{N\!\! +\! K'}\! \!\sum_{l = 1}^{N \!+ \!K'} \| L_{l, LayerCAM} \! -\!  d_{\phi}(e_{\theta}\big(L_{l, LayerCAM}) \big) \|^2_2.
	\end{split}
\end{equation}

\subsection{LayerCAM for Malicious Local Model Updates Identification}
Once the autoencoder completes the training process, the server computes the reconstruction errors between each reconstructed LayerCAM heat map and its corresponding input LayerCAM heat map and obtains the mean reconstruction error, i.e., $\forall l  \in \mathcal{K} \cup \mathcal{M},$
  \begin{equation}
  	R_l = \frac{\sum_{i = 1}^{B} \sum_{j = 1}^{H} \big| L_{l, LayerCAM}(i, j) - L^{'}_{l, LayerCAM}(i, j) \big| }{H \times B}.
  	\label{rec-error}
  \end{equation}  
  The average reconstruction error of all LayerCAM heat maps is 
  \begin{equation}
  	\overline{R} = \frac{1}{R_{l}}\sum_{l = 1}^{N+K'} R_{l}. 
  	\label{mean-error}
  \end{equation}
  A threshold $\delta$ is defined as
  \begin{equation}
  	\delta = \overline{R} + \alpha \times \sqrt{ \frac{\sum_{l = 1}^{N+K'} (R_l - \overline{R})^2 }{N+K'}},
  	\label{threshold}
  \end{equation}
  where $\alpha$ is an empirically configured efficient. 
  If the mean reconstruction error for each LayerCAM heat map exceeds the threshold $\delta$, the corresponding input of the LayerCAM heat map is considered potentially abnormal. Otherwise, it is a potential normal LayerCAM heat map because the AE learns to capture variations in normal LayerCAM heat maps during training. The AE can encounter difficulties in handling anomalies that do not conform to the learned patterns.
  
  Note that the autoencoder is optimized to minimize the reconstruction errors between the input LayerCAM heat maps and the reconstructed LayerCAM heat maps during training.
  In other words, the autoencoder learns to reconstruct normal LayerCAM heat maps. 
  It is reasonable to expect that the LayerCAM heat maps that can be reconstructed with low reconstruction errors are considered normal, while those with high reconstruction errors are potentially abnormal.

Also note that $\delta$ is time-varying with communication rounds as the LayerCAM heat maps change over the communication rounds. The  detection results are stored in a buffer:
   \begin{equation}
  	O_l^t =
  	\begin{cases}
  		1 ,      & \quad \text{if }  R_l \leq \delta; \\
  		0,  & \quad \text{if } R_l > \delta. 
  	\end{cases}
  		\label{predict-results}
  \end{equation}
  where ``1'' and ``0'' indicate a benign and a malicious local model update, respectively.

  A voting mechanism is further designed to further reduce the possibility of misclassifying a benign local model update. The key idea is that the server makes final decisions every fixed $\xi$ communication rounds 
  to determine which devices are malicious. Specifically, the local model updates $\epsilon$ out of $\xi$ communication rounds are detected as potentially malicious. They are 
  determined malicious and removed from global aggregations. 
  Finally, (\ref{contaminated-global-model}) 
  is rewritten as
  \begin{equation}
  	\mathbf{W}_g^{t+1}  =
  	\begin{cases}
  		\sum_{l = 1}^{N+K'}  O_l^t \times   \frac{D_l}{S}  \times \mathbf{W}_{l}^{t}  ,   &  \text{if } \ t \ \text{modulo} \ \xi  \neq 0; \\
  		\sum_{l=1}^{N+K'}  Y_l^t \times   \frac{D_l}{S}  \times \mathbf{W}_{l}^{t} ,  & \text{if } \ t \ \text{modulo} \ \xi  = 0, 
  	\end{cases}
  	\label{global-model-update}
  \end{equation}
  where $Y_l^t$ is given by
  \begin{equation}
  	Y_l^t =
  	\begin{cases}
  		1 ,      & \quad \text{if }  \sum_{t = \Delta \xi - \xi +1}^{\Delta \xi } O_l^t < \epsilon,  \Delta = 1,  \cdots, \lfloor \frac{T_{FL}}{\xi}  \rfloor; \\
  		0,  & \quad \text{if }\sum_{t = \Delta \xi - \xi +1}^{\Delta \xi } O_l^t \geq \epsilon,   \Delta = 1,  \cdots, \lfloor \frac{T_{FL}}{\xi}  \rfloor, 
  	\end{cases}
  	\label{indicator}
  \end{equation}
  where $\lfloor \cdot  \rfloor$ is the floor function. 
  
  The updated global model $\mathbf{W}_g^{t+1}$ is sent back to all users. The detailed pseudo-code is given in \textbf{Algorithm 1}.


\SetKwComment{Comment}{/* }{ */}
\RestyleAlgo{ruled}
\begin{algorithm}[htpb!]
	\caption{The proposed LayerCAM-AE for detecting malicious local model updates on FL}\label{alg:two}
	\SetKwInOut{Input}{Input}
	\SetKwInOut{Output}{Output}
	\Input{$N$ - The number of benign users; $T_{FL}$ - The total number of communication round; $L$  - The number of local epochs; $\eta_t$ - The learning rate of users; $\mathbf{W}^0$ - The initialized global model; $B$ - Buffer size; $B_s$ - Voting threshold; $I_0$ -  A image picked from testing dataset}
	\Output{The final global model weight $\mathbf{W}^{T_{FL}}$}
	\For{$t = 1, \cdots, T_{FL}$}{
	Server broadcasts $\mathbf{W}^t$ to all users\;
	$\mathbf{M} \gets \mathbf{0}, \mathbf{M} \in \mathbf{R}^{\xi \times (|N+K'|)} $ \;
	\For{$n \in \mathcal{N}$}{
		$\mathbf{W}_{n}^{t} \gets \underline{LocalUpdate (n, \mathbf{W}^{t})}$ \;
	}
	
	\For{$k' \in \mathcal{M}$}{
		  $ ( \mathbf{W}_{k'}^t )^{*} \gets$  Solve (\ref{attacker-obj}) with constraint (\ref{attacker-constraint}) \;	
	}

    \eIf{$ t \ \text{modulo} \ \xi  \neq 0 $}{
    	$A_l \gets$  $\mathbf{W}_{l}^{t}$ and $I_0$ perform convolution operation \;
    	$L_{l, LayerCAM} \gets$  according to (\ref{layer-importance}) and (\ref{LayerCAM-maps}) \;
    	$\mathbf{L_{LayerCAM}} \gets$ concatenate($L_{l, LayerCAM}$) \;
    	$\mathbf{L^{'}_{LayerCAM}}\gets$ Autoencoder($\mathbf{L_{LayerCAM}}$) \;
    	$O_l^{t} \gets (\ref{predict-results})$ \;
    	Store $O_l^{t}$ into $\mathbf{M}$ \;
    }{
    	$A_l \gets$  $\mathbf{W}_{l}^{t+1}$ and $I_0$ perform convolution operation \;
    	$L_{l, LayerCAM} \gets$  according to (\ref{layer-importance}) and (\ref{LayerCAM-maps}) \;
    	$\mathbf{L_{LayerCAM}} \gets$ concatenate($L_{l, LayerCAM}$) \;
    	$\mathbf{L^{'}_{LayerCAM}} \gets$ Autoencoder($\mathbf{L_{LayerCAM}}$) \;
    	$O_l^{t} \gets  (\ref{predict-results})$ \;
    	Store $O_l^{t}$ into  $\mathbf{M}$\;
    	$Y_l^{t} \gets  (\ref{indicator})$  \;
    }  
    Update global model $\mathbf{W}_g^{t+1} \gets $ (\ref{global-model-update}).
	}

	$\underline{LocalUpdate (n, \mathbf{W}^{t})}$ \;
	\For{$ \ell = 1, \cdots, L$}{
		\For{$b_n \in \mathcal{D}_n$}{
			(\ref{local-model}).
		}
	}
	\label{Algorithm 1}
\end{algorithm}

\section{Performance Evaluation} 

\subsection{Experiment Setup}

\textbf{Parameter settings.} We set the parameters of LayerCAM-AE as follows. The total number of communication rounds is $T_{FL} = 100$. For each communication round, 21 benign devices train their local models for 25 epochs using the Adam optimizer, with a batch size of 64, a learning rate of 0.0001, and a weight decay of 0. Three attackers eavesdrop on the benign local models in each communication round. 
At the server, the autoencoder trains the LayerCAM heat maps for 200 epochs using the Adam optimizer, with a hidden-layer size of 128, a learning rate of 0.001, and a weight decay of 0.00001. In the voting process, the communication round interval is $\xi=3$, and the threshold is $\epsilon=2$. The experiments were conducted on a single Zotac GeForce RTX 4090 with 24GB GDDR6X memory.  

\textbf{Datasets.} Two datasets, i.e., SVHN \cite{netzer2011reading} and CIFAR-100 \cite{krizhevsky2009learning}, are used to evaluate the performance of our proposed LayerCAM-AE.

\noindent 
\begin{itemize}
	\item   SVHN: This dataset consists of 600,000 $32 \times 32 $ color images in 10 different classes. There are 73,257 training images allocated to all benign devices for local model training, and 26,032 testing images are allocated to the server for the global model testing until the end of each communication round.
 
	\item CIFAR-100: This dataset contains 100 different classes of 60,000 $32 \times 32 $ color images. There are 50,000 training images, and 10,000 testing images are assigned to all users for training and to the server for testing. 	
\end{itemize} 

\noindent
\textbf{Benchmarks.} We consider the following state-of-the-art defense schemes, i.e., 
 \begin{itemize}
	\item  AUROR  \cite{shen2016auror} (K-means based): AUROR employs K-means to cluster the uploaded local model updates over training rounds and discards the malicious model updates, i.e., contributions from small clusters that exceed a threshold distance are considered malicious. 

    \item Multi-Krum \cite{blanchard2017machine} (Euclidean distance based): This scheme computes a score of the sum of each local model update to its neighbor's Euclidean distance. The score is the sum of its Euclidean distance from its neighbors; those with high scores as malicious model updates are excluded.
 
	\item FAA-DL \cite{faadl2022} (SVM based): This scheme utilizes an appropriate kernel function and soft margins to estimate a nonlinear decision boundary and separate the benign and malicious local model updates.
 
	\item GradCAM-AE: In this scheme, an autoencoder is applied to identify the malicious GradCAM heat maps that are generated at the server.
 
	\item LayerCAM-Krum:  In this scheme, the Krum \cite{blanchard2017machine} algorithm is applied to identify the abnormal LayerCAM heat maps that are generated at the server.
\end{itemize}

As far as the detection rates of detection methods are concerned, Recall, Precision, FPR (false positive rate), ACC (Accuracy), and F1 score are taken as performance measures, which can be obtained by relevant calculation through a confusion matrix. Moreover, we further calculate the second-level detection index AUC (area under the receiver operating characteristic curve) to measure the detection performance of the defense approaches. 

\subsection{Evaluation under IID Datasets}
\begin{figure} 
	\begin{subfigure}{.5\textwidth}
		\centering
		\includegraphics[width=1.0\linewidth]{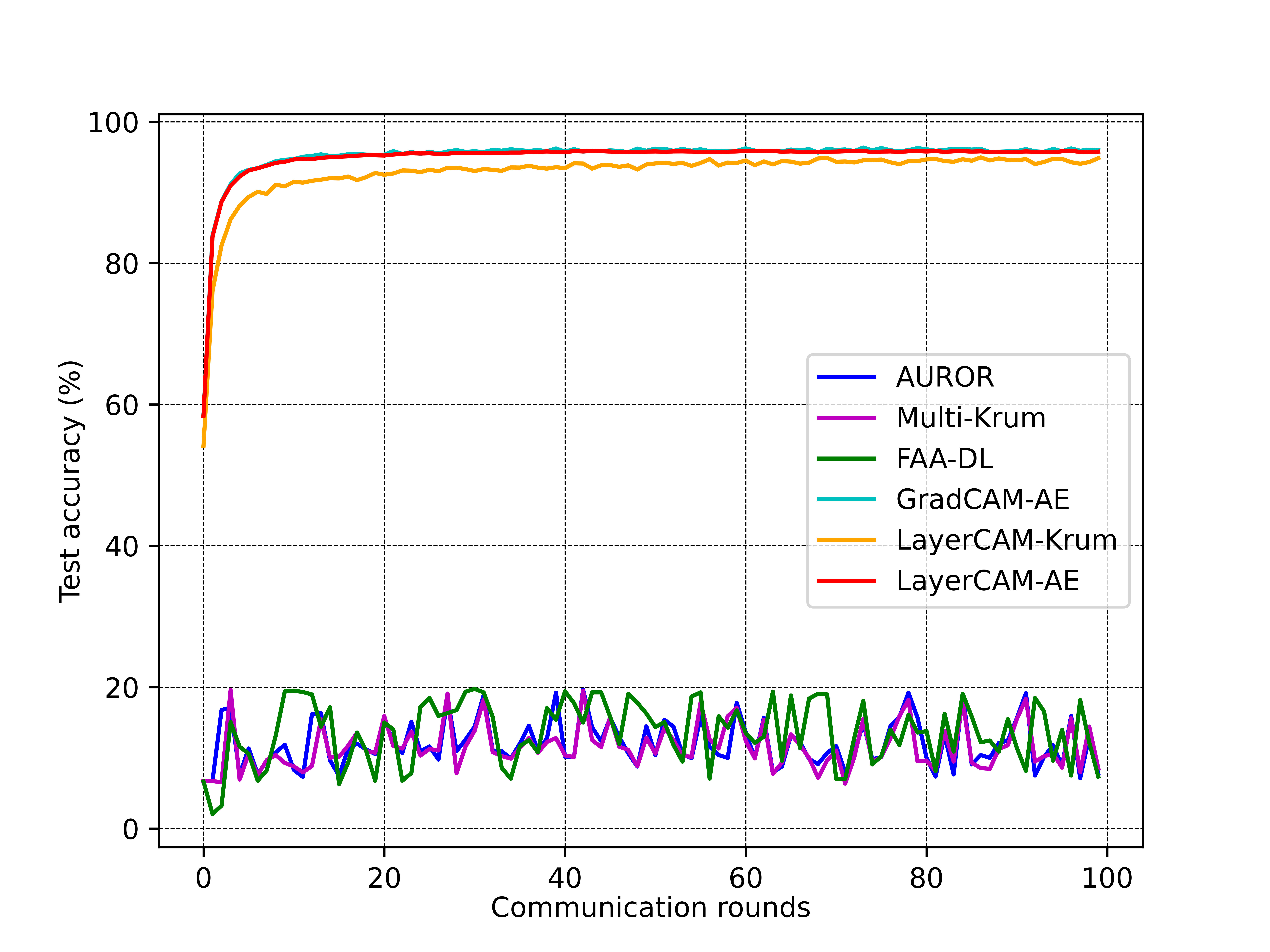}  
		\caption{SVHN}
		\label{resnet50-iid-svhn}
\end{subfigure}
\begin{subfigure}{.5\textwidth}
		\centering
		\includegraphics[width=1.0\linewidth]{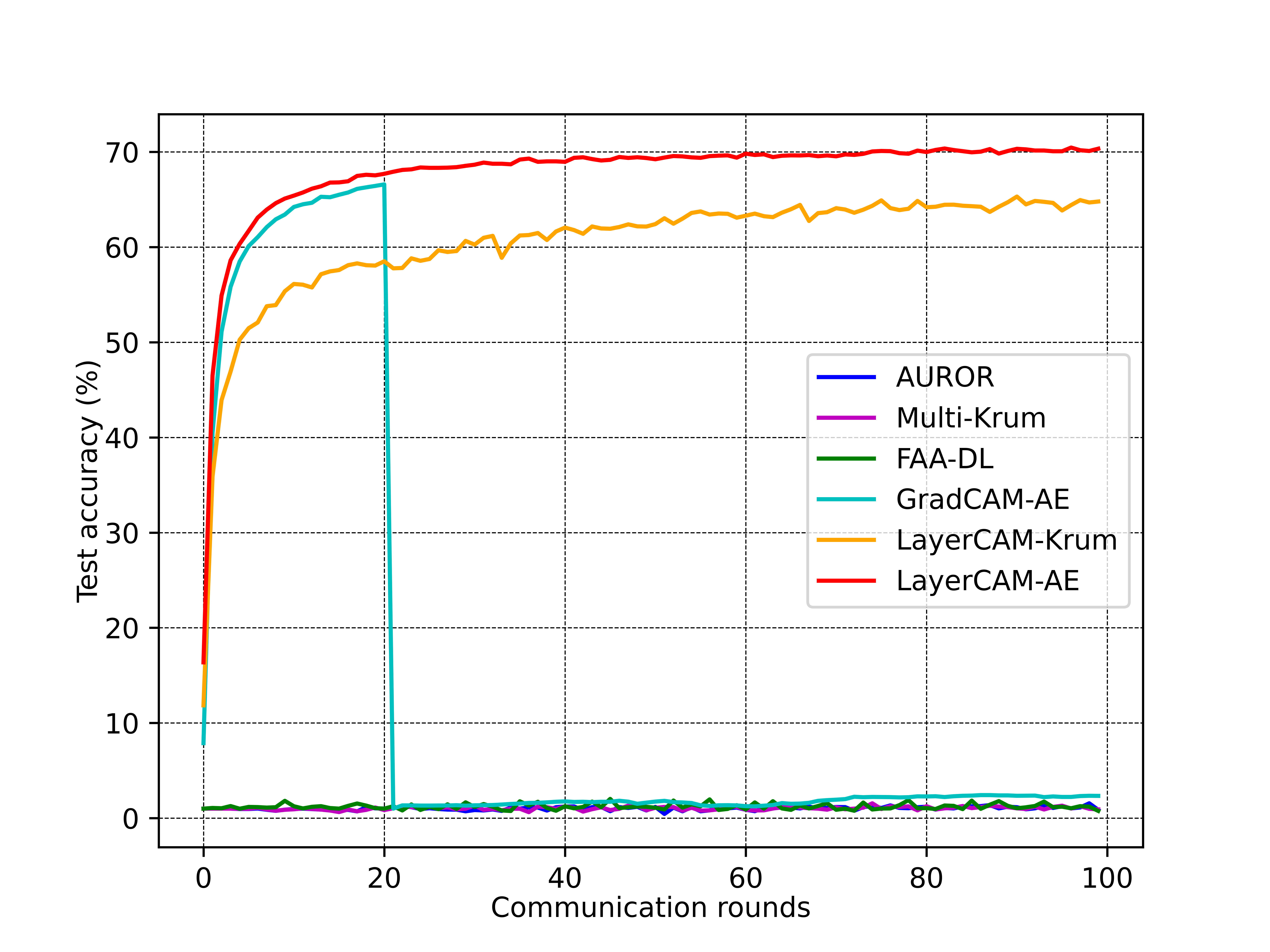}  
		\caption{CIFAR100}
		\label{resnet50-iid-cifar100}
\end{subfigure}
	\caption{ResNet-50 on IID SVHN and CIFAR100.}
	\label{resnet50-iid}
\end{figure}

Fig. \ref{resnet50-iid} compares FL test accuracy between the proposed LayerCAM-AE defense framework and the benchmarks, where the ResNet-50 model performs
image classification on SVHN and CIFAR-100 with IID
settings. 
In Fig. \ref{resnet50-iid-svhn}, LayerCAM-AE and GradCAM-AE can achieve the highest
test accuracy of the global model as the autoencoder can
accurately exclude the abnormal LayerCAM and GradCAM heat maps, respectively. As a result, more benign local model updates are involved in the FL training process. This conforms to the results of the detection rates presented in Table \ref{resnet50-iid-detection}. 
When we replace the SVHN dataset with a more complicated CIFAR-100 dataset, GradCAM-AE is unable to detect the malicious local models in the 21st communication round, causing the malicious local models to be involved in the update of the FL global model. This is because GradCAM is unable to capture fine-grained features as the dataset becomes more complex. As FL continues to train, the disappearance of the heat maps generated by benign users becomes more and more serious. In contrast, LayerCAM can capture more detailed features without compromising the detection of anomaly heat maps.

\begin{table*}
	\caption{Detection rates of \textbf{ResNet-50} on \textbf{IID} SVHN and CIFAR-100 with 3 attackers.}
	\centering
	\begin{tabular}{|c|cccccc|cccccc|}
		\hline
		\textbf{ResNet-50} & \multicolumn{6}{c|}{IID SVHN }    & \multicolumn{6}{c|}{IID CIFAR-100} \\
		\hline
		Methods & Recall &  Precision & FPR & ACC & F1 score &AUC  &Recall &  Precision  & FPR & ACC & F1 score &AUC \\
		\hline
		AUROR &  0.047 &  0.030 & 0.173 & 0.729 & 0.036 & 0.437 & 0.04 & 0.024  & 0.170 & 0.731 & 0.030 & 0.435 \\
		\hline
		Multi-Krum & 0.080 &  0.080 & 0.131 & 0.77 & 0.080 &0.474 & 0.08 &  0.08 & 0.131 & 0.77 & 0.08 & 0.474 \\
		\hline
		FAA-DL & 0.737 &  0.123 & 0.751 & 0.310 & 0.210 &0.493 & 0.763 &  0.121  & 0.783 & 0.285 & 0.208 & 0.490 \\
		\hline 
		GradCAM-AE & 1.0 &  1.0 & 0.0 & 1.0 & 1.0 & 1.0 & 1.0 & 0.948 & 0.016 & 0.986 & 0.965 & 0.992 \\
		\hline
		LayerCAM-Krum & 0.91 &  0.91 & 0.013 & 0.978 & 0.91 & 0.949 & 0.637 &  0.637  & 0.052 & 0.909 & 0.637 & 0.792 \\
		\hline
		LayerCAM-AE & \textbf{1.0} &  \textbf{1.0} & \textbf{0.0} & \textbf{1.0} & \textbf{1.0} & \textbf{1.0}  & \textbf{1.0} &  \textbf{1.0} & \textbf{0.0} & \textbf{1.0} & \textbf{1.0} & \textbf{1.0} \\
		\hline
	\end{tabular}
	\label{resnet50-iid-detection}
\end{table*}

While LayerCAM-Krum only selects an optimal local model update as the global model to train FL, it also achieves higher test accuracy (94.8\% on the IID SVHN dataset and 64.8\% on the IID CIFAR-100 dataset), compared to the benchmarks. The reason is that the data is IID, and the difference among the local model updates is insignificant.

Both AUROR and Multi-Krum directly detect millions, even many more, parameters of local model updates based on the Euclidean distance. The Euclidean distance between the crafted malicious local model updates and the benign ones is within the threshold designated by the server. This indicates that the malicious local model updates can elude the detection of the server and participate in the FL training process through multiple communication rounds, resulting in the global model being corrupted. Table \ref{resnet50-iid-detection} shows that although the ACCs of the Euclidean distance-based defense schemes are relatively high (0.729 and 0.77), the Precision is close to zero. In other words, they can correctly classify a few benign users but fail to identify malicious users. Similar performances are observed on the CIFAR-100 dataset, i.e., LayerCAM-AE still outperforms the benchmarks, as shown in the detection rates of ResNet-50 on IID CIFAR-100 in Table \ref{resnet50-iid-detection}.

\begin{figure}[htbp!]
	\begin{subfigure}{.5\textwidth}
		\centering
		\includegraphics[width=1.0\linewidth]{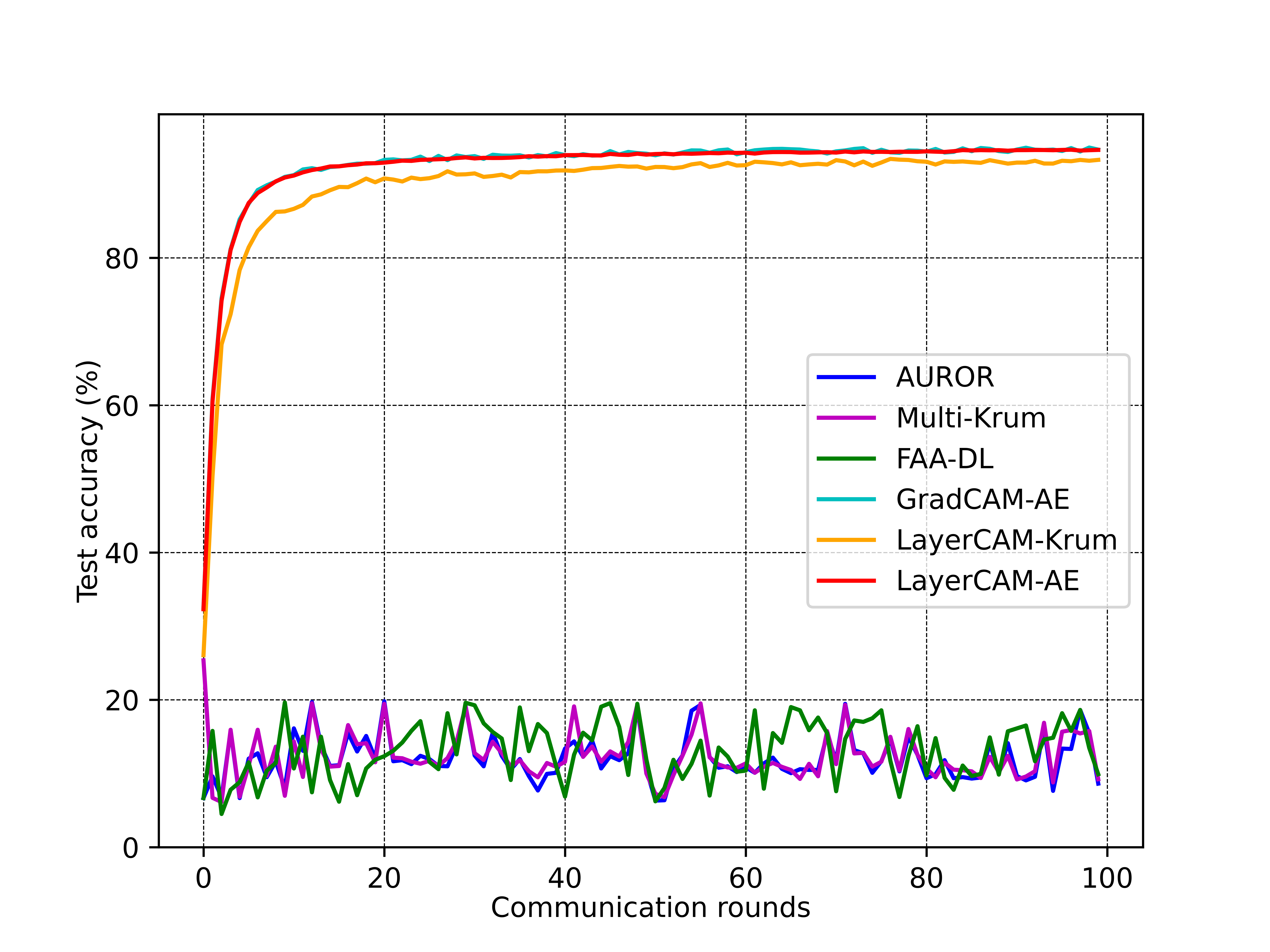}  
		\caption{SVHN}
		\label{regnet-y-800m-iid-svhn}
	\end{subfigure}
	\begin{subfigure}{.5\textwidth}
		\centering
		\includegraphics[width=1.0\linewidth]{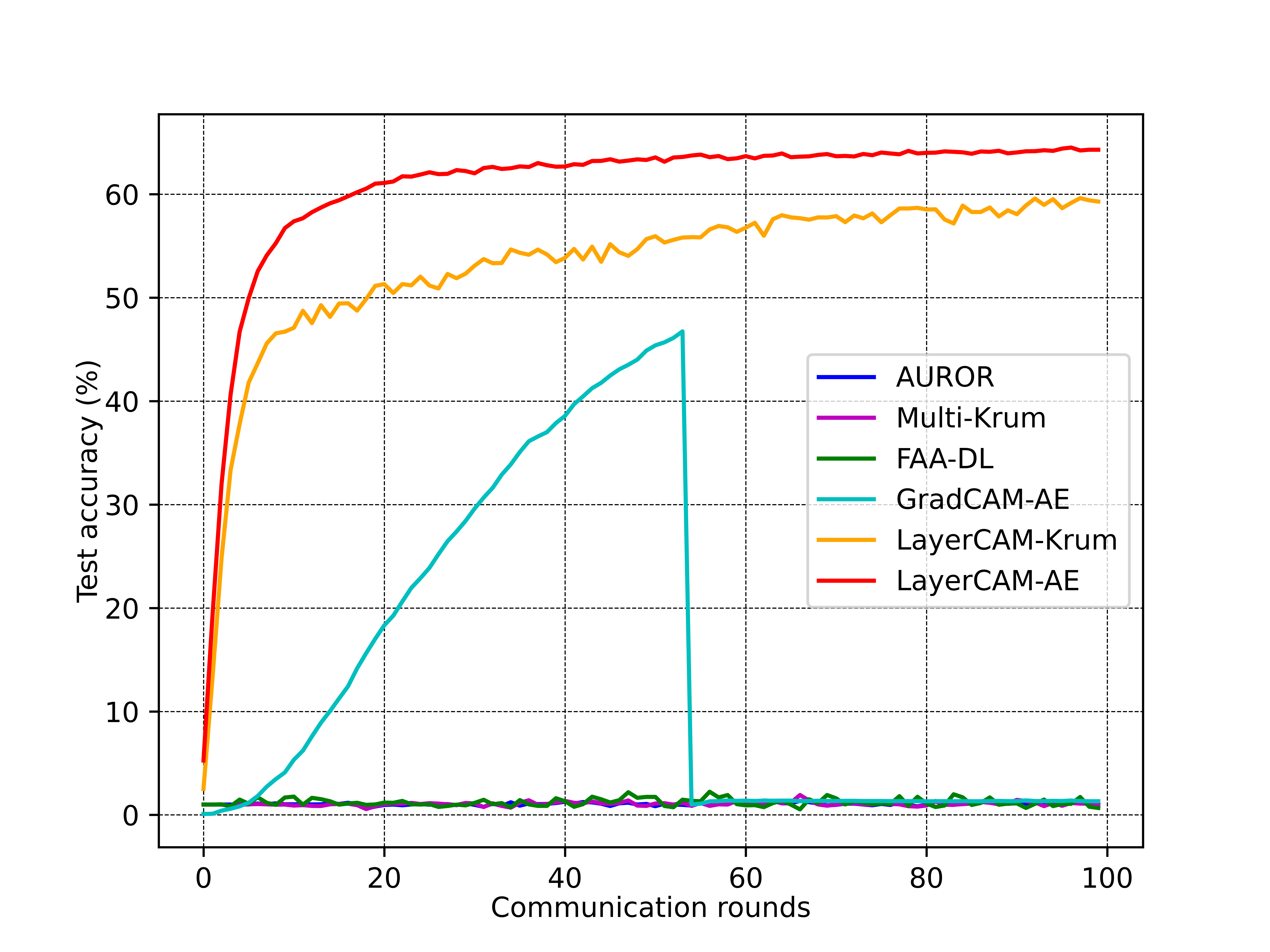}  
		\caption{CIFAR100}
		\label{regnet-y-800m-iid-cifar100}
	\end{subfigure}
	\caption{REGNETY-800MF on IID SVHN and CIFAR100.}
	\label{regnet-y-800mf-iid}
\end{figure}

FAA-DL, a one-class SVM-based method, also directly classifies the local model updates aggregated by the server as benign and malicious. However, there are two key reasons why FAA-DL fails in the experiments. On the one hand, the local model updates have the characteristics of high-dimensional feature spaces. The curse of dimensionality can lead to data sparsity, making it challenging for FAA-DL to find a suitable margin that separates benign from malicious local model updates. On the other hand, FAA-DL is sensitive to class imbalance (21 benign local model updates and three malicious local model updates); in other words, FAA-DL biases its decision boundary towards the majority class, making it struggle to detect malicious local model updates effectively. Notwithstanding, we replace the
DNN model with REGNETY-800MF. Similar performances are observed; see Fig. \ref{regnet-y-800mf-iid}. 
The detection rates of REGNETY-800MF on SVHN and CIFAR-100 with the IID settings are given in Table \ref{regnet-y-800mf-iid-detection}. This demonstrates that LayerCAM-AE is applicable to different DNN models for detecting malicious local model updates.

\begin{table*}
	\caption{Detection rates of \textbf{REGNETY-800MF} on \textbf{IID} SVHN and CIFAR-100 with 3 attackers.}
	\centering
	\begin{tabular}{|c|cccccc|cccccc|}
		\hline
		\textbf{REGNETY-800MF} & \multicolumn{6}{c|}{IID SVHN }    & \multicolumn{6}{c|}{IID CIFAR-100} \\
		\hline
		Methods & Recall &  Precision & FPR & ACC & F1 score &AUC  &Recall &  Precision  & FPR & ACC & F1 score &AUC \\
		\hline 
		AUROR &  0.020 &  - & 0.171 & 0.728 & 0.015 & 0.424 & 0.027 &  0.168  & 0.174 & 0.726 & 0.020 & 0.426\\
		\hline 
		Multi-Krum & 0.063 &  0.063 & 0.134 & 0.766 & 0.063 & 0.465 & 0.093 &  0.093 & 0.130 & 0.773 & 0.093 & 0.482 \\
		\hline 
		FAA-DL & 0.780 & 0.124 & 0.788 & 0.283 & 0.213 & 0.496  & 0.747 &  0.122  & 00.762 & 0.302 & 0.209 & 0.492 \\
		\hline 
		GradCAM-AE &  1.0 &  1.0 & 0.0 & 1.0 & 1.0 & 1.0  & 1.0 &  0.980 & 0.004 & 0.996 & 0.988 & 0.998 \\
		\hline
		LayerCAM-Krum & 0.997 &  0.997 & 0.0005 & 0.999 & 0.997 & 0.998 & 0.807 &  0.807  & 0.028 & 0.952 & 0.807 & 0.890\\
		\hline
		LayerCAM-AE & \textbf{1.0} &  \textbf{1.0} & \textbf{0.0} & \textbf{1.0} & \textbf{1.0} & \textbf{1.0} & \textbf{1.0} &  \textbf{1.0} & \textbf{0.0} & \textbf{1.0} & \textbf{1.0} & \textbf{1.0}\\
		\hline		
	\end{tabular}
	\label{regnet-y-800mf-iid-detection}
\end{table*}

\subsection{Performance under non-IID Datasets}
Compared with the IID setting, the non-IID setting is much more challenging for the server in identifying the malicious local model updates. This reason is that the local model updates tend to be more divergent under the non-IID settings; in turn,
the generated heat maps on the server are more diverse. To simulate the statistical data heterogeneity in practical FL scenarios, we adopt one of the non-IID partition schemes, i.e., latent Dirichlet allocation (LDA), which are widely used in the latest literature \cite{www1,  www2}. As a biased probability, the $\alpha$ parameter of LDA controls the distribution difference of the local training data. A larger $\alpha$ indicates a higher level of data heterogeneity under the non-IID setting among the local training data. Here, we set $\alpha = 0.5$.

\begin{figure}[htbp!]
	\begin{subfigure}{.5\textwidth}
		\centering
		\includegraphics[width=1.0\linewidth]{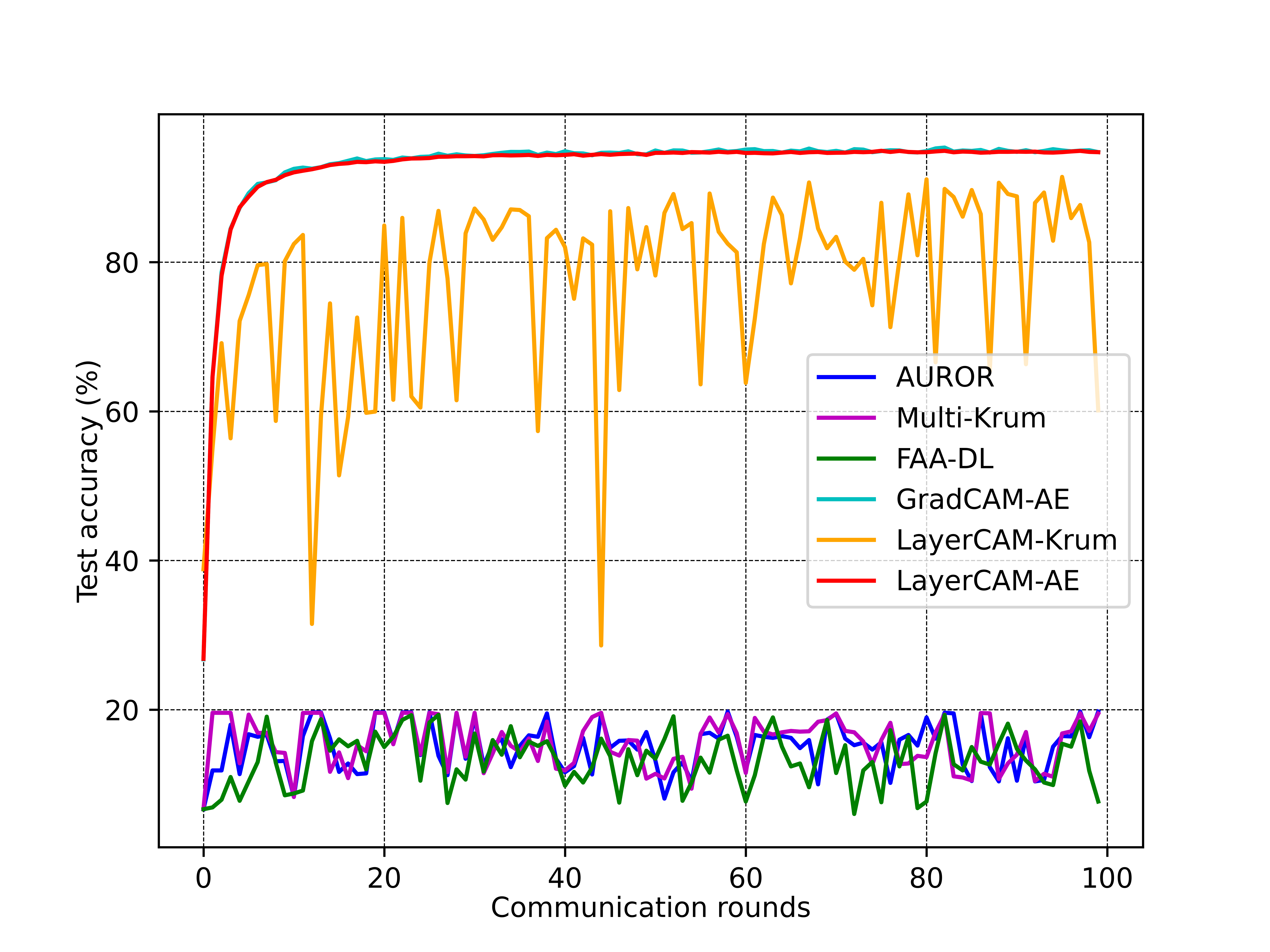}  
		\caption{SVHN}
		\label{resnet50-non-iid-svhn}
	\end{subfigure}
	\begin{subfigure}{.5\textwidth}
		\centering
		\includegraphics[width=1.0\linewidth]{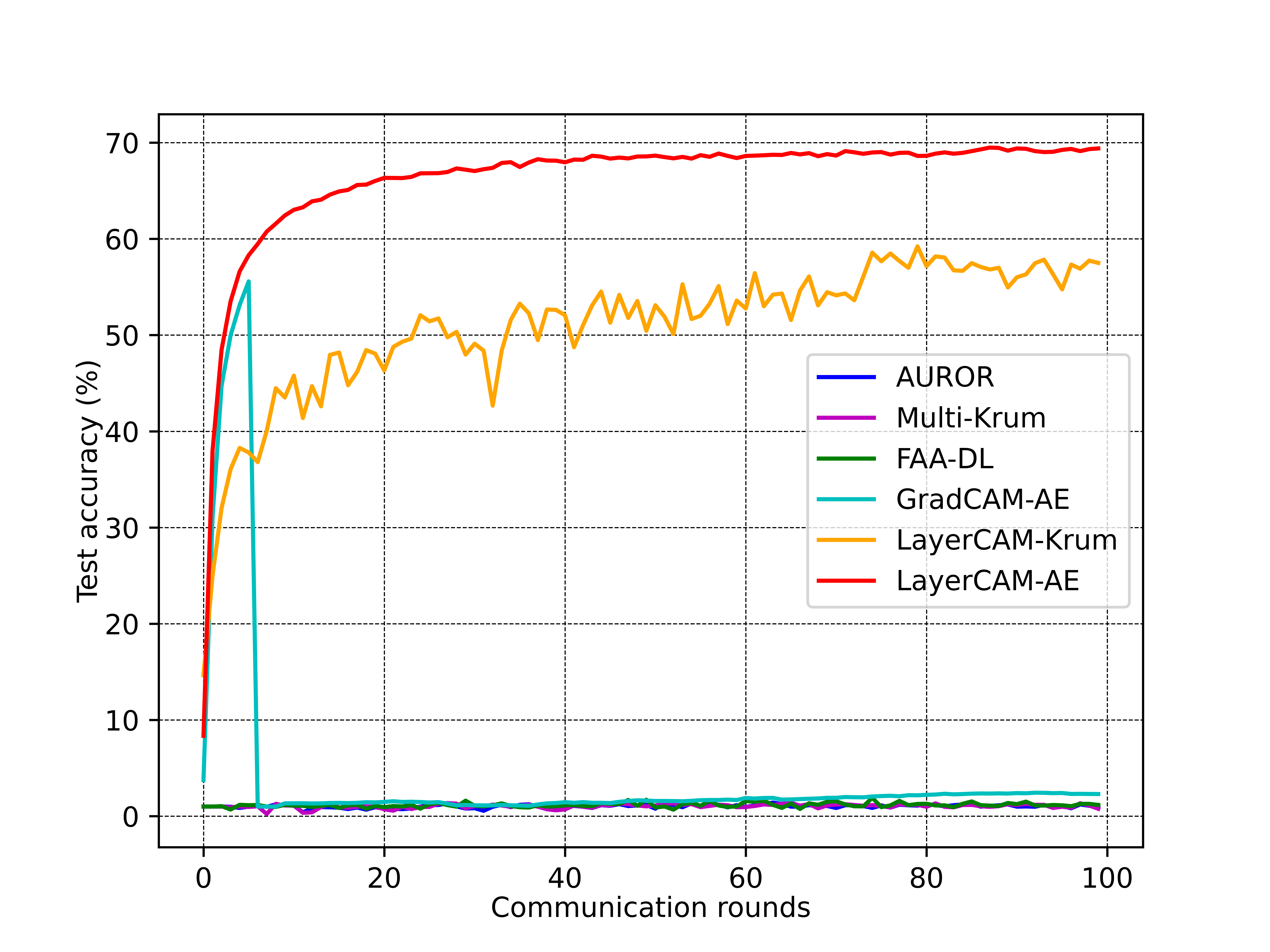}  
		\caption{CIFAR100}
		\label{resnet50-non-iid-cifar100}
	\end{subfigure}
	\caption{ResNet-50 on non-IID SVHN and CIFAR100.}
	\label{resnet50-non-iid}
\end{figure}

Fig. \ref{resnet50-non-iid} plots the test accuracy of ResNet-50 on the non-IID SVHN dataset and the non-IID CIFAR-100 dataset, demonstrating the superiority of LayerCAM-AE. In Fig. \ref{resnet50-non-iid-svhn}, LayerCAM-AE achieves the highest test accuracy under the non-IID SVHN setting. Moreover, LayerCAM-AE can quickly converge (around the tenth communication round) as it involves more benign users in FL training. This indicates that LayerCAM-AE can accurately filter malicious model updates, as can also be confirmed by the detection indicators in Table \ref{resnet50-non-iid-detection}. Although LayerCAM-Krum can prevent malicious users from participating in FL, it sacrifices accuracy and robustness, as it selects only one local model update as the global model. The more divergent the model updates, the more diverse the heat maps. LayerCAM-Krum struggles to screen malicious model updates, which coincides with the precision of 0.516 for LayerCAM-Krum on non-IID SVHN, as shown in Table \ref{resnet50-non-iid-detection}.

Compared to LayerCAM-AE, GradCAM-AE performs worse on the more complex non-IID CIFAR-100 dataset, as shown in Fig. \ref{resnet50-non-iid-cifar100}. The reason is that the local models trained by FL benign users on heterogeneous datasets cannot generate heat maps through GradCAM, resulting in GradCAM-AE's misclassification. To this end, the dataset can have an impact on GradCAM-AE. The more complex the dataset, the more difficult it is for GradCAM-AE to identify malicious users, as also confirmed in Figs. \ref{resnet50-iid-cifar100} and \ref{resnet50-non-iid-cifar100}. The remaining defense methods still undergo poor anomaly detection under the non-IID datasets, as they do on the IID datasets. The detection rates of ResNet-50 on SVHN and CIFAR-100 under the non-IID settings are given in Table \ref{resnet50-non-iid-detection}. 

\begin{table*}
	\caption{Detection rates of \textbf{ResNet-50} on \textbf{non-IID} SVHN and CIFAR-100 with 3 attackers.}
	\centering
	\begin{tabular}{|c|cccccc|cccccc|}
		\hline
		\textbf{ResNet-50} & \multicolumn{6}{c|}{non-IID SVHN }    & \multicolumn{6}{c|}{non-IID CIFAR-100} \\
		\hline
		Methods & Recall &  Precision & FPR & ACC & F1 score &AUC  &Recall &  Precision  & FPR & ACC & F1 score &AUC \\
		\hline 
		AUROR &  0.027 &  0.017 & 0.162 & 0.736 & 0.021 & 0.432 & 0.020 &  0.013  & 0.181 & 0.718 & 0.016 & 0.419 \\
		\hline 
		Multi-Krum & 0.07 &  0.07 & 0.133 & 0.768 & 0.07 & 0.469  & 0.077 &  0.077  & 0.132& 0.769 & 0.077 & 0.472\\ 
		\hline
		FAA-DL & 0.720 &  0.122 & 0.721 & 0.334 & 0.208 & 0.50  & 0.703 &  0.128  & 0.680 & 0.368 & 0.215 & 0.512 \\
		\hline 
		GradCAM-AE & 1.0 &  1.0 & 0.0 & 1.0 & 1.0 & 1.0  & 1.0 &  0.95 & 0.010 & 0.992 & 0.971 & 0.999 \\
		\hline
		LayerCAM-Krum & 0.517 &  0.517 & 0.069 & 0.879 & 0.516 & 0.724  & 0.337 &  0.337  & 0.095 & 0.834 & 0.337 & 0.621 \\
		\hline
		LayerCAM-AE & \textbf{1.0} &  \textbf{1.0} & \textbf{0.0} & \textbf{1.0} & \textbf{1.0} & \textbf{1.0} & \textbf{1.0} &  \textbf{1.0} & \textbf{0.0} & \textbf{1.0} & \textbf{1.0} & \textbf{1.0}\\
		\hline		
	\end{tabular}
	\label{resnet50-non-iid-detection}
\end{table*}

\begin{figure}[htbp!]
	\begin{subfigure}{.5\textwidth}
		\centering
		\includegraphics[width=1.0\linewidth]{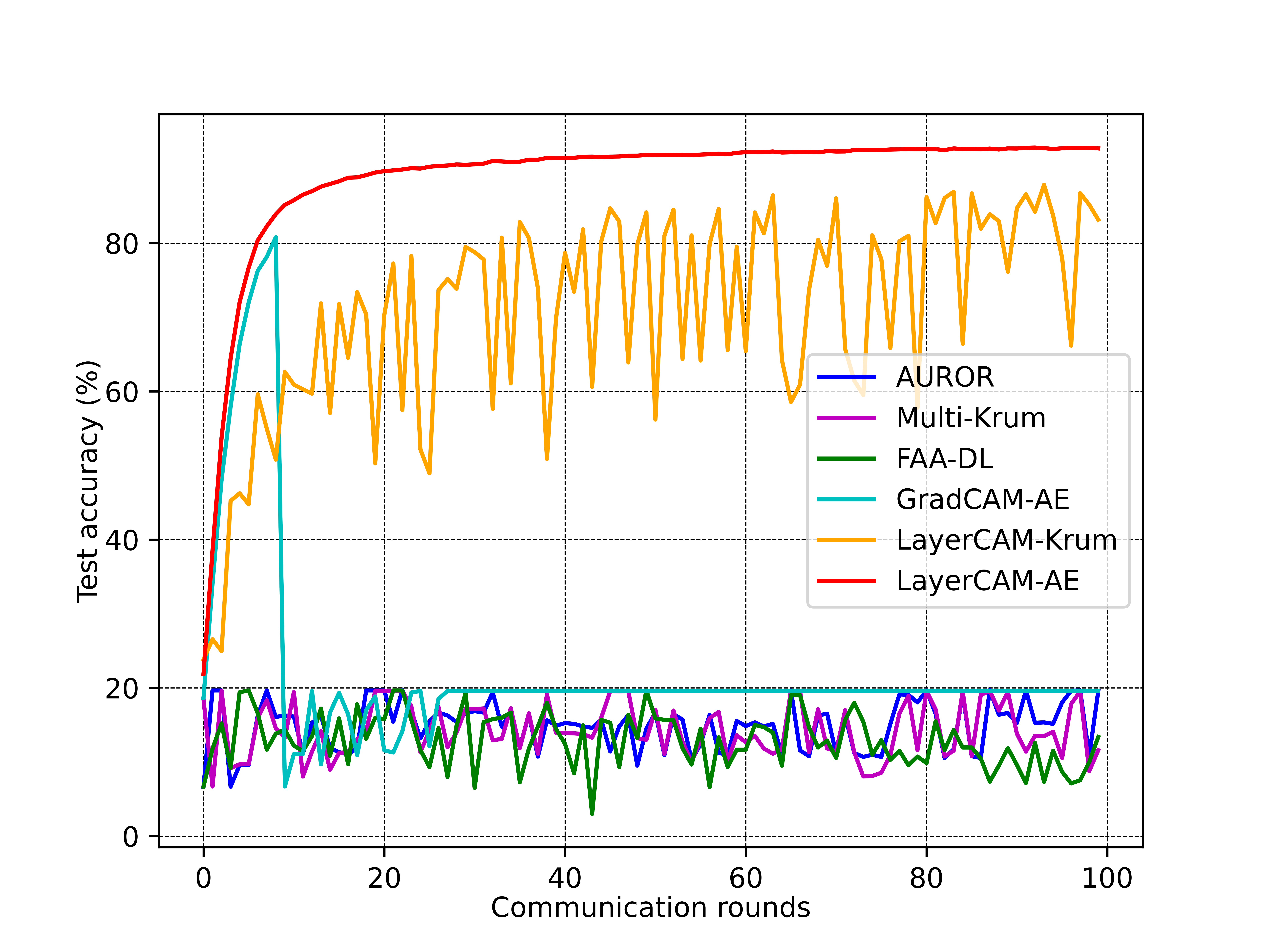}  
		\caption{SVHN}
		\label{regnet-y-800m-non-iid-svhn}
	\end{subfigure}
	\begin{subfigure}{.5\textwidth}
		\centering
		\includegraphics[width=1.0\linewidth]{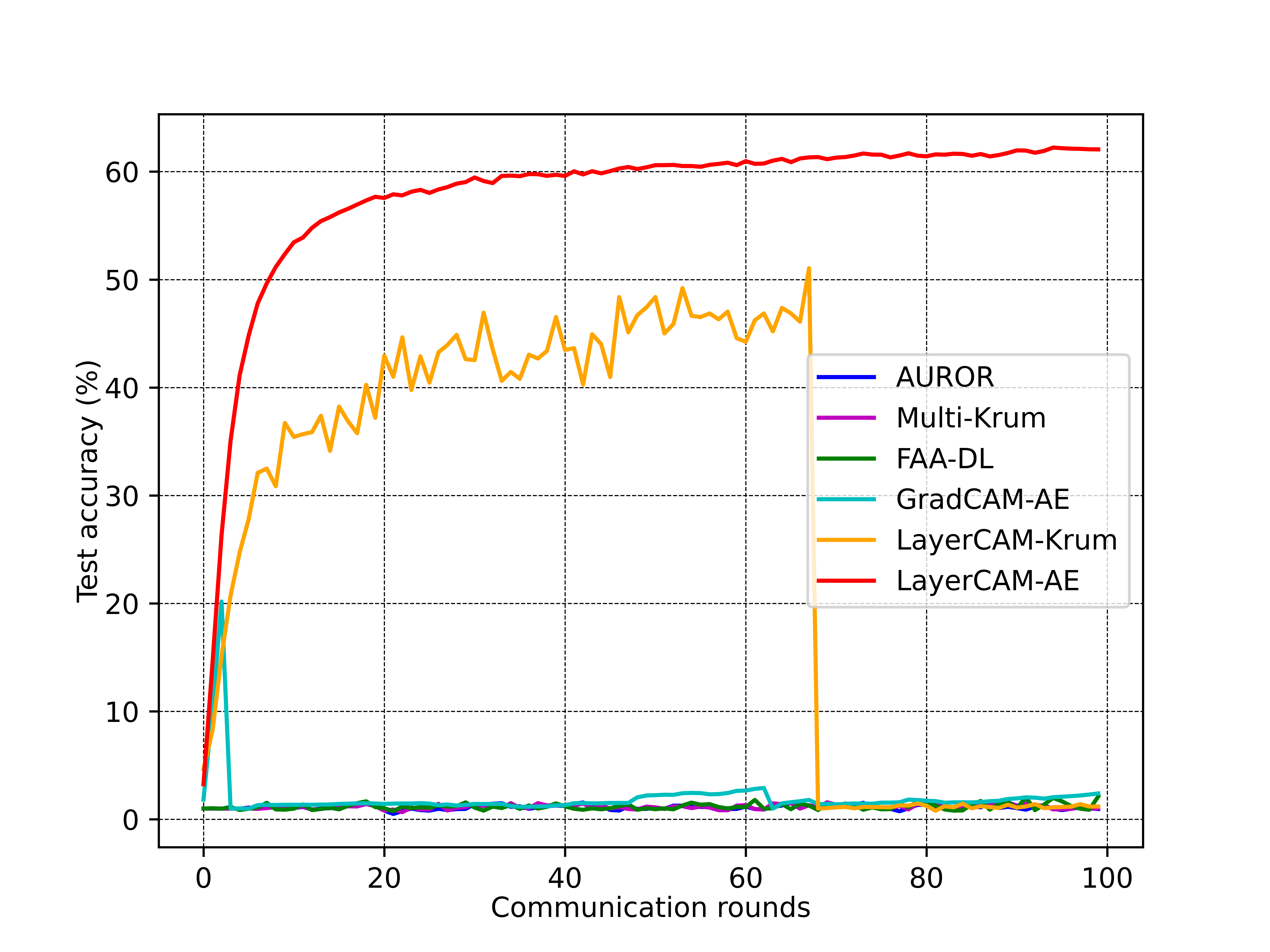}  
		\caption{CIFAR100}
		\label{regnet-y-800m-non-iid-cifar100}
	\end{subfigure}
	\caption{REGNETY-800MF on non-IID SVHN and CIFAR100.}
	\label{regnet-y-800mf-non-iid}
\end{figure}

We replace the ResNet-50 model with the REGNETY-800MF model. The trend of the test accuracy of the defense methods with the communication rounds is consistent with the observation in Fig. \ref{resnet50-non-iid-svhn}, except for GradCAM-AE, as shown in Fig. \ref{regnet-y-800m-non-iid-svhn}.  The reason is that the performance of GradCAM may vary depending on the architecture of the neural network analyzed. It may not be as effective for models with complex architectures, such as attention-based models, where the relationships between features are more intricate. 

As contrasted with GradCAM-AE, LayerCAM-AE is designed to be compatible with various network architectures, including both traditional CNNs and more complex architectures such as residual networks (ResNets) or REGNETY-800MF. This adaptability allows LayerCAM-AE to be applied across a wide range of models without modification. In other words, LayerCAM-AE is unaffected by the structure of DNNs, which is also evident in Figs. \ref{resnet50-non-iid-svhn} and \ref{resnet50-non-iid-cifar100}.  In Fig. \ref{regnet-y-800m-non-iid-cifar100}, LayerCAM-Krum mistakenly selects the malicious local model update as the global model update in the 68th communication round, causing the test accuracy of the global model to drop sharply. The detection rates of REGNETY-800MF on SVHN and CIFAR-100 under the non-IID settings are given in Table \ref{regnet-y-800mf-non-iid-detection}.

\begin{table*}
	\caption{Detection rates of  \textbf{REGNETY-800MF} on \textbf{non-IID} SVHN and CIFAR-100 with 3 attackers.}
	\centering
	\begin{tabular}{|c|cccccc|cccccc|}
		\hline
		\textbf{REGNETY-800MF} & \multicolumn{6}{c|}{non-IID SVHN }    & \multicolumn{6}{c|}{non-IID CIFAR-100} \\
		\hline
		Methods & Recall &  Precision & FPR & ACC & F1 score &AUC  &Recall &  Precision  & FPR & ACC & F1 score &AUC \\
		\hline 
		AUROR &  0.057 &  0.042 & 0.153 & 0.748 & 0.047 & 0.452 & 0.013 &  0.008  & 0.158 & 0.738 & 0.01 & 0.427 \\
		\hline 
		Multi-Krum & 0.073 &  0.073 & 0.132 & 0.768 & 0.073 & 0.470 & 0.1 &  0.1 & 0.129 & 0.775 & 0.1 & 0.486\\
		\hline
		FAA-DL & 0.743 &  0.127 & 0.732 & 0.327 & 0.215 & 0.505  & 0.727 &  0.123 & 0.730 & 0.327 & 0.210 & 0.498\\ 
		\hline
		GradCAM-AE & 0.848 &  - & 0.004 & 0.977 & 0.844 & 0.922  & 0.828 &  0.828  & 0.02 & 0.974 & 0.820 & 0.917 \\
		\hline 
		LayerCAM-Krum & 0.547 &  0.547 & 0.065 & 0.887 & 0.547 & 0.741  & 0.95 &  0.95 & 0.007 & 0.987 & 0.95 & 0.971\\
		\hline
		LayerCAM-AE & \textbf{1.0} &  \textbf{1.0} & \textbf{0.0} & \textbf{1.0} & \textbf{1.0} & \textbf{1.0} & \textbf{1.0} &  \textbf{1.0} & \textbf{0.0} & \textbf{1.0} & \textbf{1.0} & \textbf{1.0}\\
		\hline		
	\end{tabular}
	\label{regnet-y-800mf-non-iid-detection}
\end{table*}

\begin{figure*}[htbp!]
	\begin{subfigure}{1.0\textwidth}
		\centering
		\includegraphics[width=1.0\linewidth]{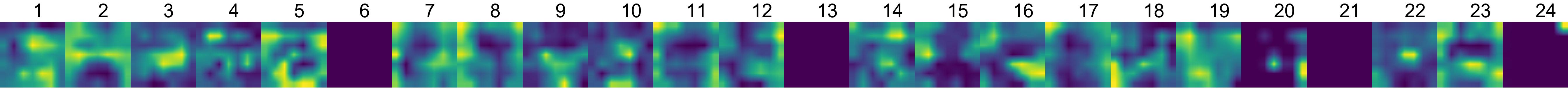}  
		\caption{Generated heat maps by LayerCAM (original)}
		\label{fig:sub-first}
	\end{subfigure}
    \vfill
	\begin{subfigure}{1.0\textwidth}
		\centering
		\includegraphics[width=1.0\linewidth]{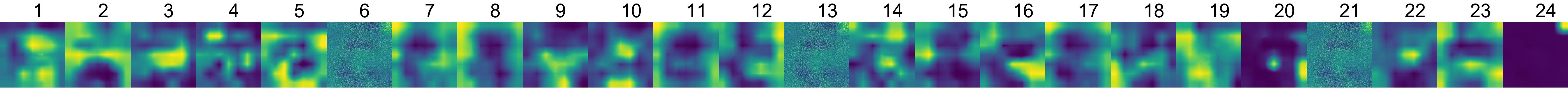}  
		\caption{Reconstructed heat maps by autoencoder}
		\label{fig:sub-second}
	\end{subfigure}
	\caption{Comparison between generated heat maps by LayerCAM and reconstructed heat maps by the autoencoder under REGNETY-800MF on non-IID CIFAR-100 dataset with $\alpha = 0.1$.}
	\label{reconstruct-heat-maps}
\end{figure*}

The considerable distinction is delineated in Table \ref{reconstruction-errors}, where the server iteratively computes the average reconstruction errors across 24 users over 100 communication rounds. Fig. \ref{reconstruct-heat-maps} showcases the original LayerCAM heat maps alongside the reconstructed heat maps generated by the autoencoder under the REGNETY-800MF configuration in the non-IID CIFAR-100 setting with $\alpha = 0.1$. Remarkably, it becomes evident that the reconstruction errors in Table \ref{reconstruction-errors} for malicious users 6, 13, and 21—measuring 0.214 each—are notably higher than those of the other benign users.

\begin{table}
\caption{Reconstruction errors for REGNETY-800MF on non-IID CIFAR-100 dataset with $\alpha = 0.1$}
 \centering
  \begin{tabular}{|c|c|}
    \hline
    Index of users &  reconstruction errors \\
    \hline
    1 & $ 0.007 \pm 0.005$  \\ 
     \hline
    2 &  $ 0.006 \pm 0.004$  \\ 
     \hline
    3 & $ 0.007 \pm 0.004$   \\ 
     \hline
     4 & $ 0.007 \pm 0.006$  \\ 
     \hline
    5 & $ 0.007 \pm 0.004$  \\ 
     \hline
    \textbf{6} & $ \textbf{0.214} \pm 0.004$  \\ 
     \hline
     7 & $ 0.007 \pm 0.004$  \\ 
     \hline
     8 & $ 0.007 \pm 0.004$  \\ 
     \hline
     9 & $ 0.007 \pm 0.005$  \\ 
     \hline
    10 & $ 0.007 \pm 0.005$   \\ 
     \hline
     11 & $ 0.007 \pm 0.004$  \\ 
     \hline
     12 & $ 0.007 \pm 0.005$  \\ 
     \hline
     \textbf{13} & $ \textbf{0.214} \pm 0.004$  \\ 
     \hline
    14 & $ 0.007 \pm 0.004$  \\ 
     \hline
     15 & $ 0.006 \pm 0.004$  \\ 
     \hline
     16 & $ 0.007 \pm 0.005$  \\ 
     \hline
       17 & $ 0.007 \pm 0.006$   \\ 
     \hline
    18 & $ 0.007 \pm 0.004$  \\ 
     \hline
     18 & $ 0.007 \pm 0.005$  \\ 
     \hline
     20 & $ 0.007 \pm 0.005$  \\ 
     \hline
       \textbf{21} & $ \textbf{0.214} \pm 0.004$  \\ 
     \hline
    22 & $ 0.006 \pm 0.004$  \\ 
     \hline
     23 & $ 0.007 \pm 0.005$   \\ 
     \hline
     24 & $ 0.007 \pm 0.005$   \\ 
     \hline
  \end{tabular}
  \label{reconstruction-errors}
\end{table}

\subsection{Scalability}
\begin{figure}[htbp!]
	\begin{subfigure}{0.5\textwidth}
		\centering
		\includegraphics[width=1.0\linewidth]{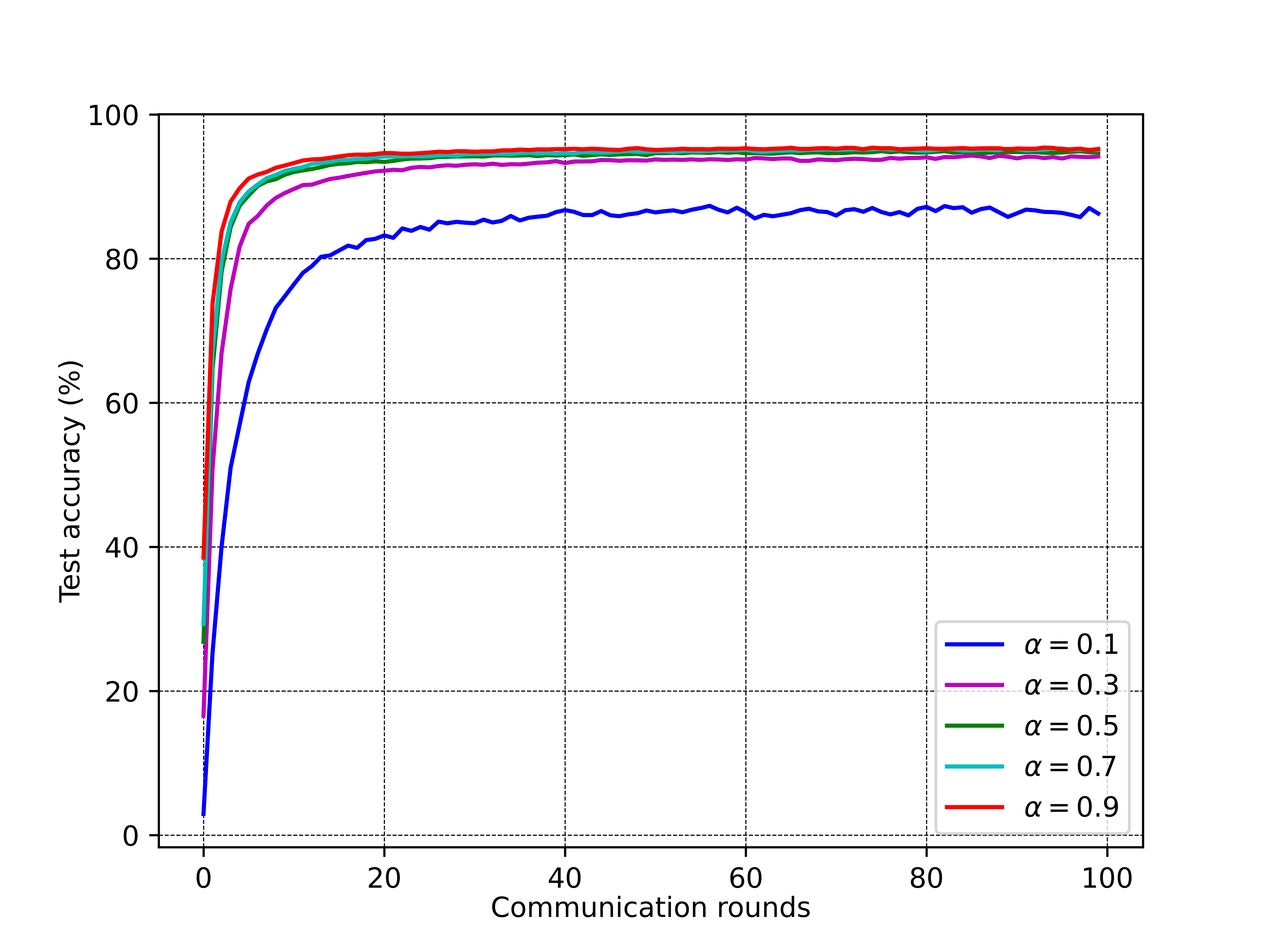}  
		\caption{SVHN}
	\end{subfigure}
    \vfill
	\begin{subfigure}{0.5\textwidth}
		\centering
		\includegraphics[width=1.0\linewidth]{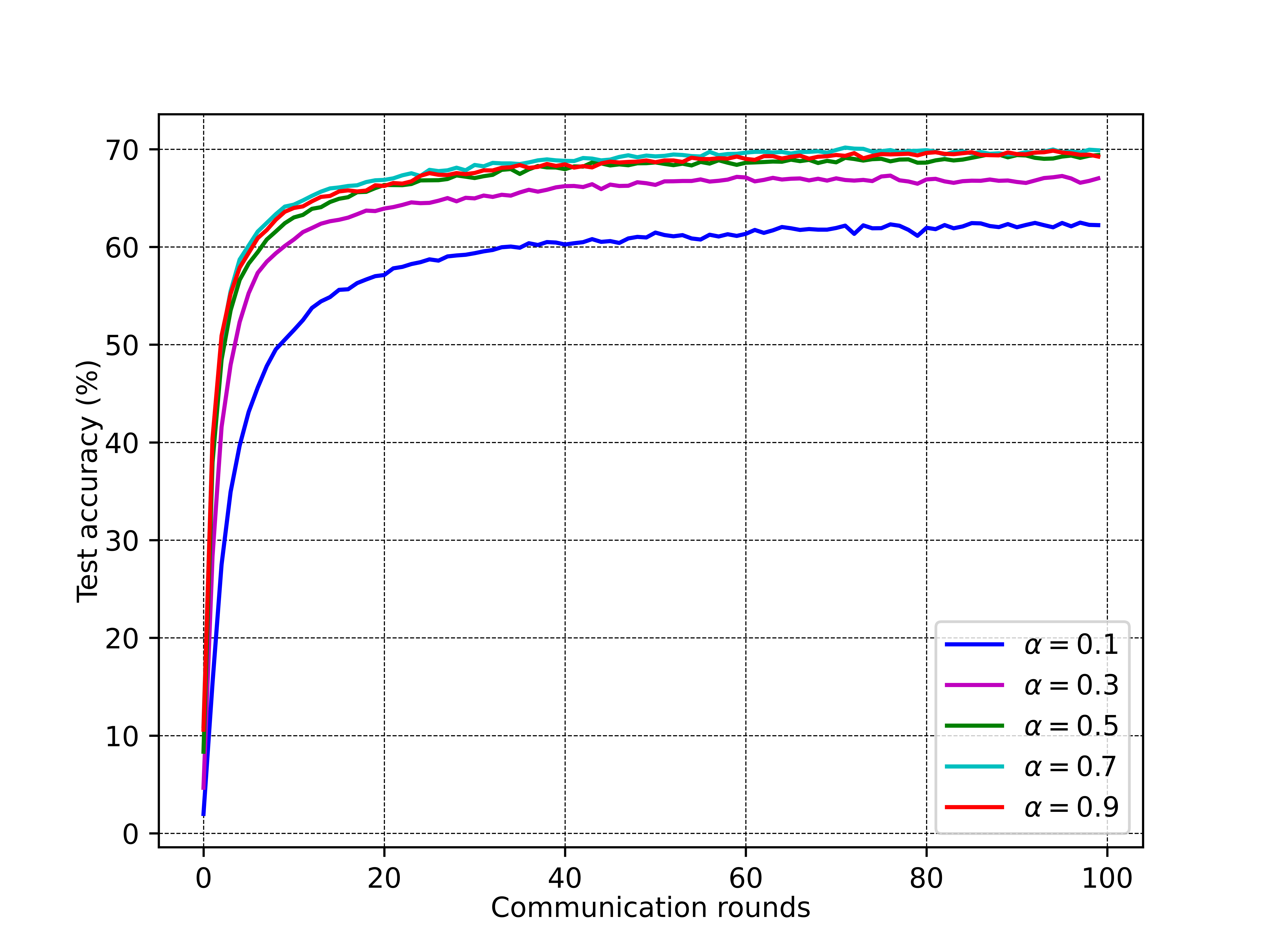}  
		\caption{CIFAR-100}
	\end{subfigure}
	\caption{The test accuracy of FL global model varies with the degree of heterogeneity ($\alpha$)  with regard to the ResNet-50 on SVHN and CIFAR-100 dataset.}
	\label{resnet50-alpha}
\end{figure}

\begin{figure}[htbp!]
	\begin{subfigure}{0.5\textwidth}
		\centering
		\includegraphics[width=1.0\linewidth]{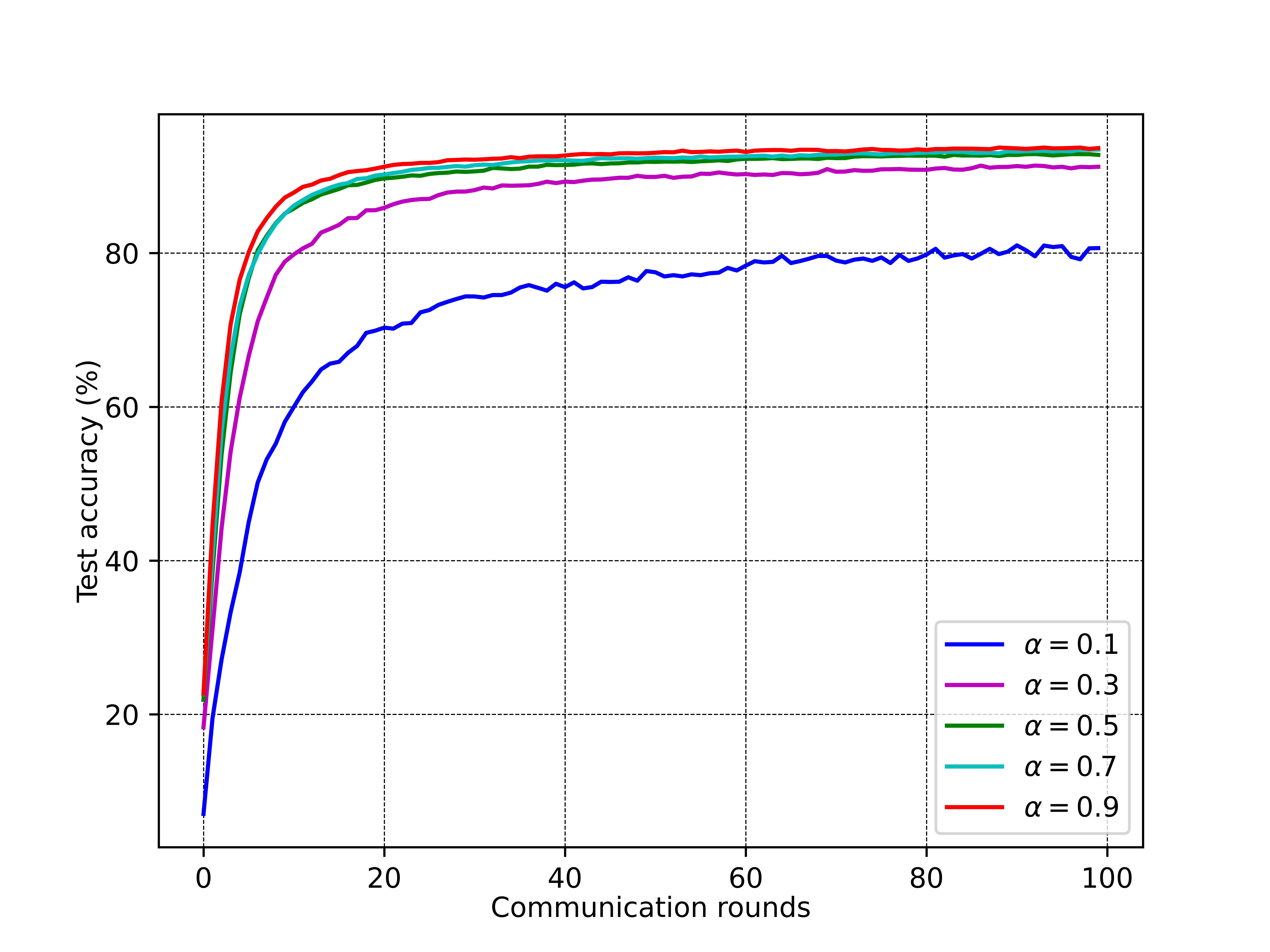}  
		\caption{SVHN}
	\end{subfigure}
    \vfill
	\begin{subfigure}{0.5\textwidth}
		\centering
		\includegraphics[width=1.0\linewidth]{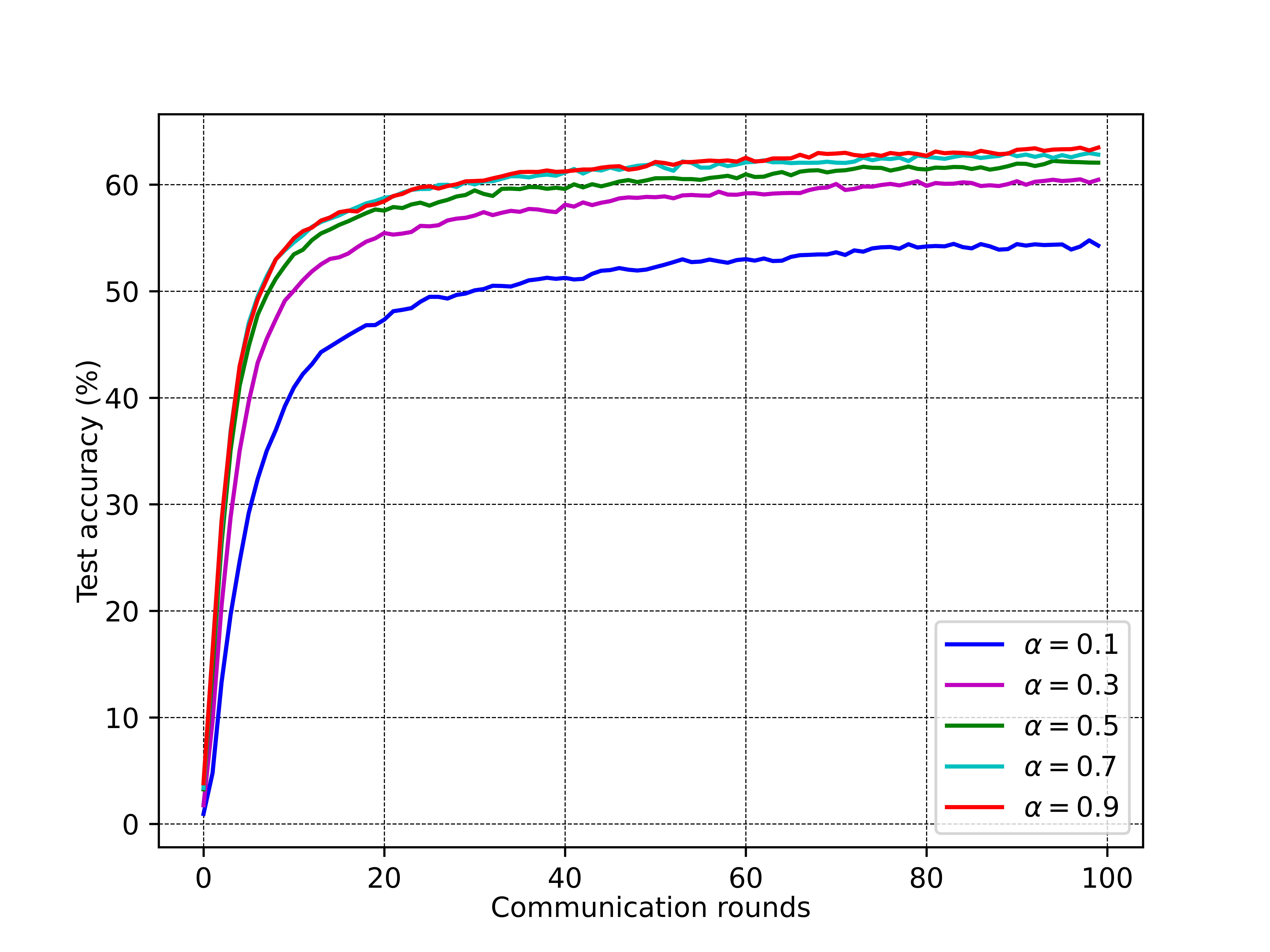}  
		\caption{CIFAR-100}
	\end{subfigure}
	\caption{The test accuracy of FL global model varies with the degree of heterogeneity ($\alpha$)  with regard to the REGNETY-800MF on SVHN and CIFAR-100 dataset.}
	\label{regnet_y_800mf-alpha}
\end{figure}

Figs. \ref{resnet50-alpha} and \ref{regnet_y_800mf-alpha} exhibit the global model's test accuracy of ResNet-50 and REGNETY-800MF under different degrees of heterogeneity ($\alpha = 0.1$, $\alpha = 0.3$, $\alpha = 0.5$, $\alpha = 0.7$ and $\alpha = 0.9$ ) on the SVHN and CIFAR-100 datasets, respectively. The experimental results indicate that LayerCAM-AE is nearly unaffected by the structures of DNNs and the degree of dataset heterogeneity.

\begin{figure}[htbp!]
  \centering
  \includegraphics[width=0.5\textwidth]{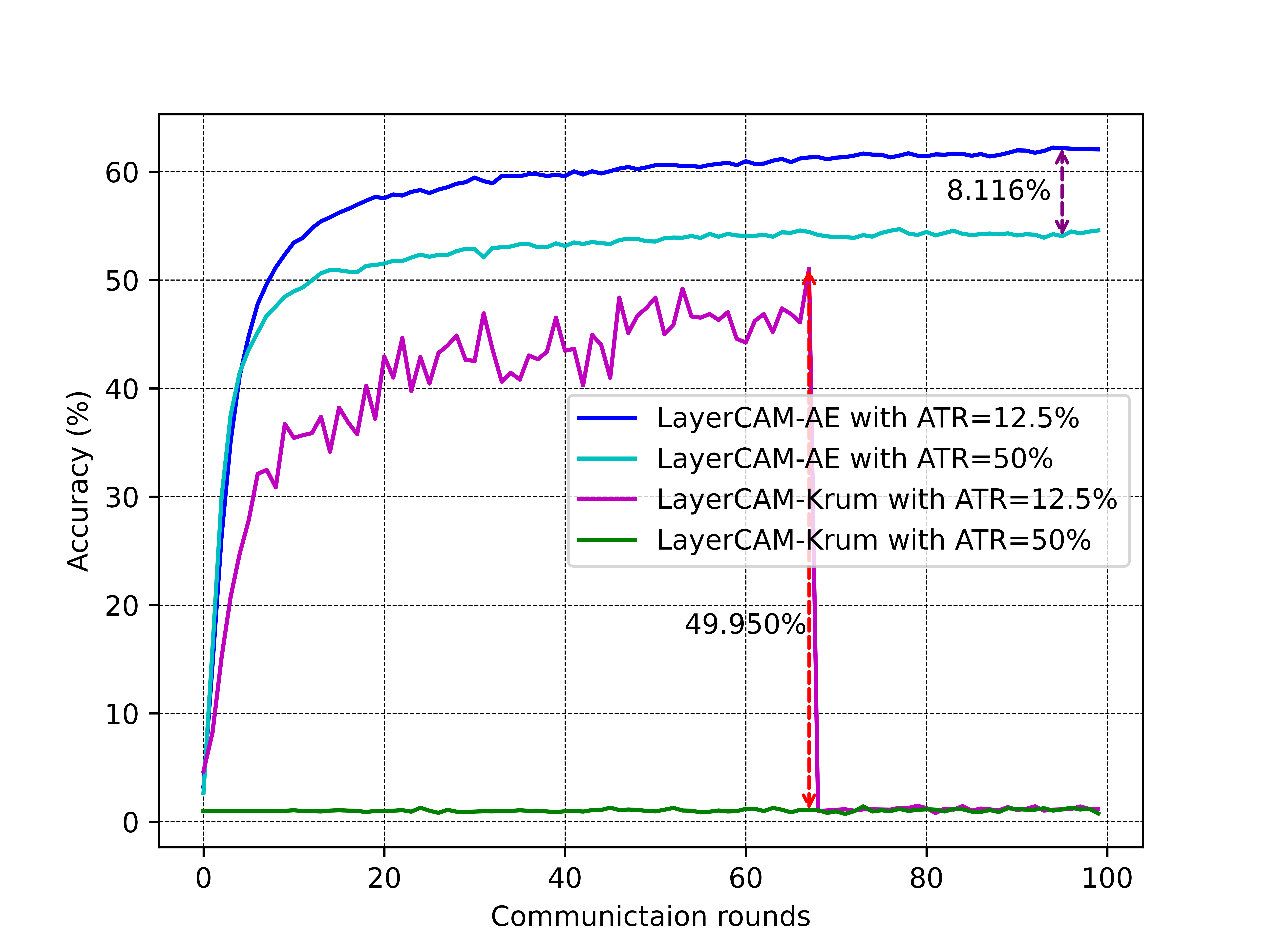}
  \caption{Scalability test.}
\label{scalability}
\end{figure}

To further evaluate the scalability of LayerCAM-AE, we define the attack rate as the ratio of the number of attackers to the total number of local users, i.e., $\rm{ATR} = \frac{\# attackers }{\# \rm{all \ 
		users} } \times 100\% = \frac{K'}{N + K'}  \times 100\%$. 
As depicted in Figure \ref{scalability}, LayerCAM-AE and LayerCAM-Krum were evaluated under non-IID conditions using the CIFAR-100 dataset with an imbalance factor ($\alpha = 0.5$), facing scenarios with 3 and 12 attackers. Given 24 local users, the ATR of LayerCAM-Krum from increases 12.5\% to 50\% as LayerCAM-Krum struggles to detect the attackers, the maximum gap of test accuracy is 49.950\% (in the 67th communication round) much higher than 8.116\% that is the maximum gap of test accuracy of LayerCAM-AE in the 96th communication round. This highlights the enhanced scalability and reliability of LayerCAM-AE.

\subsection{Ablation Studies}
\begin{figure*}[htbp!]
	\begin{subfigure}{1.0\textwidth}
		\centering
		\includegraphics[width=1.0\linewidth]{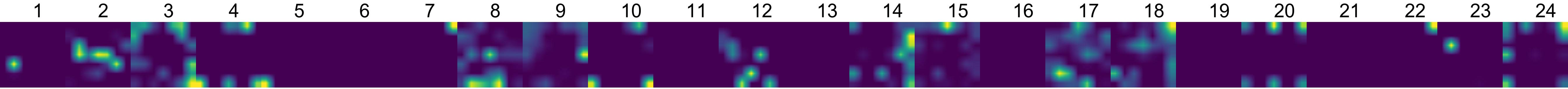}  
		\caption{heat maps generated by \textbf{GradCAM} of REGNETY-800MF with non-IID ($\alpha =0.1$) CIFAR-100 dataset in $6$-th communication round}
		\label{fig:gradcam-heatmaps}
	\end{subfigure}
	\vfill
	\begin{subfigure}{1.0\textwidth}
		\centering
		\includegraphics[width=1.0\linewidth]{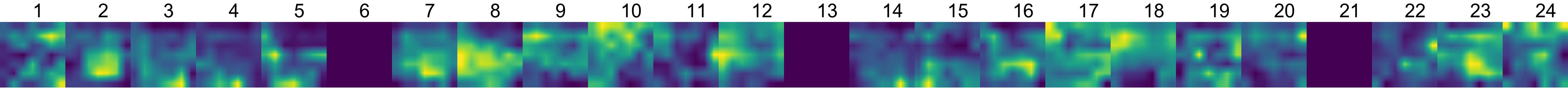}  
		\caption{heat maps generated by \textbf{LayerCAM} of REGNETY-800MF with non-IID ($\alpha =0.1$) CIFAR-100 dataset in $6$-th communication round}
		\label{fig:layercam-heatmaps}
	\end{subfigure}
	\caption{Comparison between generated heat maps by GradCAM and LayerCAM on REGNETY-800MF with non-IID ($\alpha = 0.1$) CIFAR-100 dataset  in $6$-th communication round.}
	\label{fig:heatmaps-com}
\end{figure*}

Figs. \ref{fig:gradcam-heatmaps} and \ref{fig:layercam-heatmaps} show the heat maps generated by GradCAM and LayerCAM on REGNETY-800MF on the non-IID ($\alpha = 0.1$) CIFAR-100 dataset in the sixth communication round, respectively. Obviously, it is observed from the heat maps that LayerCAM is better at capturing image features than GradCAM. While users 5, 11, 16, and 19 are benign, the test images and their local model updates via the GradCAM process fail to generate heat maps and are mistakenly considered malicious.  However, LayerCAM-AE does not produce misclassification because it accurately captures the image features. This can also be evidenced by the test accuracy of LayerCAM-AE in Figs. \ref{resnet50-non-iid-cifar100}, \ref{regnet-y-800m-non-iid-svhn}, and \ref{regnet-y-800m-non-iid-cifar100}. To this end, LayerCAM is essential to the LayerCAM-AE architecture.

We replace the autoencoder in LayerCAM-AE with Krum, and plot the test accuracy of LayerCAM-Krum in Figs. \ref{resnet50-non-iid-svhn}, \ref{resnet50-non-iid-cifar100}, \ref{regnet-y-800m-non-iid-svhn}, and \ref{regnet-y-800m-non-iid-cifar100}. The test accuracy fluctuates significantly and fails to converge. There are even malicious local model updates involved in FL global model updates. This is because, in the absence of the autoencoder remapping of the LayerCAM heat maps, LayerCAM-Krum is unable to highlight the hidden features of the heat maps and improve the heat map distinguishability. For this reason, the autoencoder is indispensable for the LayerCAM-AE architecture.

Last but not least, we further evaluate the overhead of LayerCAM-AE. This is conducted by testing the running time of LayerCAM-AE in each communication round. The running time of LayerCAM and the training time of the autoencoder are about 0.83 seconds and 12.4 seconds, respectively. Furthermore, a local user does not require any additional overhead since LayerCAM-AE is deployed on the server side. LayerCAM-AE is user-friendly.

\section{Related Work}

This section reviews the literature on defense models against poisoning attacks on FL. 

\textbf{Euclidean distance-based defense approaches.} Several existing approaches, based on Euclidean distance, have been developed to detect poisoning attacks in federated learning (FL) scenarios. Within this framework, methods like Krum \cite{blanchard2017machine, zhiyuan2022} or Multi-Krum \cite{blanchard2017machine, zhiyuan2022} calculate a score for each local model update, which represents the sum of its Euclidean distance from neighboring updates.  Multi-Krum identifies local model updates with high scores as potentially malicious and excludes them. In contrast, the Trimmed-mean methodology \cite{yin2018byzantine} is a coordinate-wise aggregation approach that aggregates each coordinate of the local model update independently. For each coordinate, the values of corresponding coordinates in the users’ updates are sorted. The largest and smallest $k$ values are then removed, and the trimmed mean calculates the average of the remaining values as the corresponding coordinate of the aggregated model update. To mitigate poisoning attacks involving numerous malicious users, FLDetector \cite{zhang2022fldetector} has been developed to predict a client's model update in each communication round based on historical updates. If the received local model update from a user consistently differs from the predicted update across multiple communication rounds, it is flagged as malicious. Another defense strategy against poisoning attacks on FL \cite{cao2022flcert} involves categorizing users into distinct groups and training a global model for each group using an existing FL aggregation rule. A majority vote mechanism, based on the global models of all groups, is then used to determine if a test input has been tampered with by the attacker.

\textbf{Machine learning-based defense.}  
to identify malicious local users while ensuring the generation of accurate models, a statistical mechanism known as AUROR has been introduced \cite{shen2016auror}. AUROR is founded on the observation that the primary model features from the majority of honest users demonstrate a consistent distribution, whereas those from malicious users exhibit an aberrant distribution. AUROR utilizes K-means to cluster uploaded local model updates across training communication rounds and eliminates malicious local model updates, i.e., contributions from small clusters that surpass a predefined distance threshold are flagged as malicious. 

In another approach, the authors of \cite{faadl2022} presented the Federated Anomaly Analytics enhanced Distributed Learning (FAA-DL) framework, which is a lightweight, unsupervised anomaly detection method based on support vector machine (SVM). FAA-DL employs an appropriate kernel function and soft margins to estimate a nonlinear decision boundary, effectively segregating benign and malicious local model updates. Given the criticality of detecting network attacks, the authors of \cite{d2mif} developed the FL framework, an integrated isolation forest algorithm, to identify and filter malicious local model updates before global model aggregation. They argue that the leaf nodes representing the malicious model are closer to the root, facilitating their detection. Deep reinforcement learning methods are employed to dynamically adjust the detection threshold for identifying malicious local model updates. In combating generative adversarial network (GAN) attacks, the authors of \cite{chen2021} devised a system that isolates local model update parameters from all users, preventing attackers from setting up GANs to carry out attacks.

The current defense mechanisms against malicious local model updates, particularly those based on Euclidean distances, face significant challenges related to the ``curse of dimensionality" in the context of deep neural networks (DNNs). This is particularly true considering that local model update parameters can encompass millions or even billions of parameters, including both weights and biases. Within high-dimensional spaces, Euclidean distances may inadvertently exaggerate the distances between model updates, leading to reduced effectiveness in detecting malicious updates. Machine learning-based detection mechanisms often heavily rely on fine-tuning hyperparameters and setting thresholds with precision. However, recent experimental findings~\cite{Yue2021, Yue2022} suggest that this approach may be prone to ineffective anomaly detection.

On the contrary, the proposed LayerCAM-AE framework introduces a novel approach to detecting malicious local model updates in federated learning (FL) by leveraging LayerCAM, thus departing from conventional Euclidean distance metrics. The LayerCAM-AE defense mechanism against poisoning attacks in FL is adept at transforming the high-dimensional local model parameters—often numbering in the millions—into low-dimensional and fine-grained LayerCAM heat maps. Moreover, an autoencoder is devised alongside LayerCAM to improve the distinguishability of the heat maps, thus boosting the accuracy in identifying abnormal heat maps generated by malicious local model updates.

\section{Conclusion}
In this paper, we proposed LayerCAM-AE, a new LayerCAM and autoencoder-assisted defense mechanism against model poisoning attacks in FL. The proposed defense mechanism is designed to process the local model updates by employing a specific image from the test dataset to generate LayerCAM heat maps. The autoencoder was extended to enhance the visibility of hidden features within the LayerCAM heat maps. Moreover, the voting algorithm was developed to consistently filter out transient malicious model updates, thereby reducing the likelihood of erroneously identifying malicious local models. Experimental results showed that the proposed LayerCAM-AE offers superior detection rates (Recall: 1.0, Precision: 1.0, FPR: 0.0, Accuracy: 1.0, F1 score: 1.0, AUC:
1.0) and FL test accuracy (69\%) with RestNet-50 on non-IID CIFAR-100 dataset and significantly surpasses contemporary defense approaches across various settings. 


\bibliographystyle{ieeetr}
\bibliography{references}
\end{document}